\documentclass[twocolumn]{aastex63}

\usepackage{amsmath}
\usepackage{booktabs}
\usepackage{threeparttable}
\usepackage{tabularx}

\submitjournal{ApJL}

\shorttitle{Bursty Star Formation and High-$z$ Galaxy LFs}
\shortauthors{Sun et al.}
\graphicspath{{./}{figures/}}

\begin{document}

\defcitealias{BL_2011}{BL11}

\title{Bursty Star Formation Naturally Explains the Abundance of Bright Galaxies at Cosmic Dawn}

\correspondingauthor{Guochao Sun} \email{guochao.sun@northwestern.edu}
\author{Guochao Sun}
\affiliation{CIERA and Department of Physics and Astronomy, Northwestern University, 1800 Sherman Ave, Evanston, IL 60201, USA}
\author{Claude-Andr\'{e} Faucher-Gigu\`{e}re}
\affiliation{CIERA and Department of Physics and Astronomy, Northwestern University, 1800 Sherman Ave, Evanston, IL 60201, USA}
\author{Christopher C. Hayward}
\affiliation{Center for Computational Astrophysics, Flatiron Institute, 162 Fifth Avenue, New York, NY 10010, USA}
\author{Xuejian Shen}
\affiliation{TAPIR, California Institute of Technology, Pasadena, CA 91125, USA}
\affiliation{Department of Physics \& Kavli Institute for Astrophysics and Space Research, Massachusetts Institute of Technology, Cambridge, MA 02139, USA}
\author{Andrew Wetzel}
\affiliation{Department of Physics \& Astronomy, University of California, Davis, CA 95616, USA}
\author{Rachel K. Cochrane}
\affiliation{Center for Computational Astrophysics, Flatiron Institute, 162 Fifth Avenue, New York, NY 10010, USA}

\begin{abstract}
Recent discoveries of a significant population of bright galaxies at cosmic dawn $\left(z \gtrsim 10\right)$ have enabled critical tests of cosmological galaxy formation models. In particular, the bright end of the galaxy UV luminosity function (UVLF) appears higher than predicted by many models. Using approximately 25,000 galaxy snapshots at $8 \leq z \leq 12$ in a suite of FIRE-2 cosmological ``zoom-in'' simulations from the Feedback in Realistic Environments (FIRE) project, we show that the observed abundance of UV-bright galaxies at cosmic dawn is reproduced in these simulations with a multi-channel implementation of standard stellar feedback processes, without any fine-tuning. Notably, we find no need to invoke previously suggested modifications such as a non-standard cosmology, a top-heavy stellar initial mass function, or a strongly enhanced star formation efficiency. We contrast the UVLFs predicted by bursty star formation in these original simulations to those derived from star formation histories (SFHs) smoothed over prescribed timescales (e.g., 100\,Myr). The comparison demonstrates that the strongly time-variable SFHs predicted by the FIRE simulations play a key role in correctly reproducing the observed, bright-end UVLFs at cosmic dawn: the bursty SFHs induce order-or-magnitude changes in the abundance of UV-bright ($M_\mathrm{UV} \lesssim -20$) galaxies at $z \gtrsim 10$. The predicted bright-end UVLFs are consistent with both the spectroscopically confirmed population and the photometrically selected candidates. We also find good agreement between the predicted and observationally inferred integrated UV luminosity densities, which evolve more weakly with redshift in FIRE than suggested by some other models. 
\end{abstract}


\section{Introduction}

For the first time, the \textit{James Webb Space Telescope (JWST)} has unlocked the door to a population-level analysis of galaxies well into the era of cosmic dawn \cite[for a review of key high-redshift science themes of \textit{JWST}, see][]{Robertson_2022ARA&A}. Following its discovery of an unexpectedly high abundance of UV-bright, massive galaxy candidates at redshift $z\gtrsim10$ \cite[e.g.,][]{Finkelstein_2022, Naidu_2022, Donnan_2023, Harikane_2023ApJS, Yan_2023}, there is a long list of intriguing questions to be answered about how to interpret these observations. What is the true nature (redshift, mass, metallicity, age, etc.) of these bright galaxies? If they are truly massive galaxies at cosmic dawn, what makes it possible for them to have formed so early? Are these observations in significant tension with the standard $\Lambda$CDM cosmological model? Observational and theoretical investigations into these questions are being actively pursued in a large body of recent literature from different perspectives, including the purity of high-$z$ candidates \citep{Naidu_2022arXiv, ArrabalHaro_2023, Curtis-Lake_2023NatAs, FM_2023, Zavala_2023}, the physics of star formation in high-$z$ galaxies \citep{Dekel_2023, MF_2023, Robertson_2023NatAs, Qin_2023, SippleLidz_2023, Trinca_2023}, the implications of high-$z$ observations for the cosmological model \citep{Boylan-Kolchin_2023, Hassan_2023, Melia_2023}, and so forth. 

While spectroscopic follow-up studies for many of the galaxy candidates are still ongoing, conservative lower limits on the bright end of UV luminosity function (UVLF) and the integrated UV luminosity density at $z \gtrsim 10$ derived from the existing, spectroscopically confirmed samples have already suggested milder redshift evolution towards $z>10$ than expected by many theoretical models \cite[e.g.,][]{Harikane_2023arXiv}. Such a higher-than-expected abundance of bright galaxies based on secure redshifts is consistent with earlier studies based on photometrically selected samples, thus calling for a re-examination of the theoretical landscape of galaxy formation at cosmic dawn\footnote{Some recent studies found that galaxies with properties similar to observed ones could be reproduced in simulations \citep[e.g.,][]{Keller_2023, McCaffrey_2023}. However, these studies did not directly model the UVLF and compare it with available JWST measurements.}. Several physical mechanisms have been considered to explain a high abundance of bright galaxies at high redshifts. For example, a higher star formation efficiency (SFE) resulting from less efficient feedback regulation could boost the UV-bright galaxy abundance by forming more stars per unit baryon \citep{Dekel_2023, Harikane_2023ApJS}, whereas a more top-heavy initial mass function (IMF) of the stellar population could similarly lead to more bright galaxies by creating more UV photons per unit stellar mass formed \citep{Inayoshi_2022, Yung_2023}. 
A conspiracy between the redshift evolution of dust attenuation and the abundance of massive halos at high $z$ could also potentially allow the bright-end UVLF and UV luminosity density to evolve relatively mildly \citep{Ferrara_2023, MF_2023}, although such a coincidence would not by itself explain the correct absolute abundance of bright galaxies.
A number of studies have also examined the possibility that the high abundance of early massive galaxies implies physics beyond the standard $\Lambda$CDM cosmology, such as a modified primordial power spectrum (\citealt{HiranoYoshida_2023}; \citealt{Padmanabhan_2023}; \citealt{ParashariLaha_2023}; though see \citealt{Sabti_2023}), primordial non-Gaussianity \citep{Biagetti_2023}, or alternative dark matter models \citep{Chang_2023, DayalGiri_2023, Gong_2023}. 

Another promising avenue to elevate the abundance of bright galaxies is the strong time variability (``burstiness'') of star formation. 
In recent years, several different galaxy formation simulations have predicted that the star formation rate (SFR) is highly time-variable in low-mass galaxies \citep[e.g.,][]{Hopkins_2014, Dominguez_2015, Muratov_2015, Sparre_2017, PallottiniFerrara_2023}. 
The prediction of bursty star formation appears generic to codes that resolve the clustering of supernovae in the interstellar medium \citep{Hu_2023}. 
The simulations predict that bursty star formation is especially common in low-mass galaxies, likely due to the shallow potential wells which allow clumpy, cold inflows and outflows to drive repeated inflow-star formation-outflow cycles \citep[][]{Stern_2021_ICV, Gurvich_2023, Byrne_2023, Hopkins_2023}. 
Since low-mass galaxies dominate at high redshift, we expect the implications of bursty star formation on the UVLF to be particularly important in this regime \citep[e.g.,][]{FM_2022}. 
Indeed, evidence for an increased level of bursty star formation has emerged from recent \textit{JWST} observations of cosmic dawn galaxies \cite[e.g.,][]{Dressler_2023, Endsley_2023, Looser_2023QG, Looser_2023}. As pointed out in recent theoretical studies \citep[][]{Mason_2023, MF_2023, Shen_2023, Munoz_2023}, 
an increased level of UV variability sourced by bursty star formation can give rise to more UV-bright galaxies due to the Eddington bias, which flattens the bright end of the UVLF. In this case, the observed UVLFs at $z\gtrsim10$ could potentially be explained by bursty star formation combined with ``normal'' SFE and production efficiency of UV photons. While bursty star formation can in principle enhance the abundance of bright galaxies, it remains to be shown whether the enhancement is sufficient to reproduce the observed bright-end of the UVLF in a self-consistent galaxy formation model, such as those provided by hydrodynamic simulations. 

In this Letter, we use a suite of cosmological ``zoom-in'' simulations from the Feedback in Realistic Environments (FIRE) project\footnote{See the FIRE project website: \url{http://fire.northwestern.edu}.} to investigate the effects of bursty star formation on the UVLF at $8 \leq z \leq 12$. 
In these simulations, the SFR variability arises self-consistently from the modeling of standard stellar feedback processes.  
It is noteworthy that these simulations --- generated before the launch of \textit{JWST} --- were in particular not in any way tuned to match recent observations. Moreover, the simulations use exactly the same FIRE-2 code \citep{Hopkins_2018} that has been used to evolve large sets of simulated galaxies all the way to $z=0$ and demonstrated to produce broadly realistic galaxy properties down to the present time \citep[e.g.][and references therein]{Wetzel_2023}. 
This is in contrast with many other simulations of cosmic dawn galaxies, in which the simulations are stopped at high redshift and for which we therefore do not know how the feedback model performs at lower redshifts. 
We show that the FIRE-2 simulations produce an excellent match to the UVLF recently measured by \textit{JWST} during cosmic dawn, and that the time variability of star formation plays an important role in explaining the observations at the bright end. 
These results constitute an important test of the feedback model and highlight the importance of considering the variability of star formation when modeling high-$z$ observations. 

Throughout the Letter, we adopt a flat $\Lambda$CDM cosmology consistent with \citet{Planck_2018}, and all magnitudes are quoted in the AB system \citep{OG_1983}. 

\section{Simulations and Analysis Methods} \label{sec:sim}

\subsection{The Simulations}

In this Letter, we analyze the same set of simulations as recently studied by \cite{Sun_2023}, which is a subset of the \textit{High-Redshift} suite \citep{Ma_2018_size, Ma_2018_lf, Ma_2019} of the FIRE-2 cosmological zoom-in simulations \citep{Hopkins_2018}. The FIRE-2 simulations use the GIZMO code with its meshless-finite mass (MFM) hydro solver \citep[][]{Hopkins_2015_GIZMO}, and include multiple channels of stellar feedback to regulate star formation. 
Star formation occurs in dense molecular gas ($n_\mathrm{H} > 1000\,\mathrm{cm^3}$) that is self-gravitating and self-shielding. 
The stellar feedback mechanisms implemented include: (1) energy, momentum, mass, and metal injection from core collapse and Type Ia supernovae and winds from OB and AGB stars, (2) photoionization and photoelectric heating, and (3) radiation pressure. 
A redshift-dependent but homogeneous ionizing background is also included following \cite{FG09}.\footnote{The version of the ionizing background used in these simulations reionizes the universe at $z_{\rm reion}\approx10$, which is earlier than the mid-point of reionization of $z_{\rm reion}\approx8$ favored by more recent observational constraints \citep[e.g.,][]{Planck_2018, FG20}. However, our main results focus on the bright end of the UVLF, which arises from relatively massive halos, whereas the suppression of galaxy formation due to heating by the ionizing background primarily affects low-mass halos \citep[$M_{\rm h} \lesssim 10^{9}$ $M_{\odot}$; e.g.][]{Gnedin_2000, Noh_McQuinn_2014}. Moreover, an earlier reionization redshift implies that in the present simulations, galaxy formation is suppressed starting earlier in the small halos, so adopting a more up-to-date reionization model would (if anything) enhance the predicted UV luminosity density. Similar arguments apply to other IGM heating processes.}  
The baryonic (dark matter) mass resolution of the set of simulations considered in this work is $m_\mathrm{b} = 7 \times 10^3\,M_{\odot}$ ($m_{\rm DM} = 4 \times 10^4\,M_{\odot}$), except for the simulations z5m11a and z5m11b, which have $m_\mathrm{b} \approx 1 \times 10^3\,M_{\odot}$ ($m_{\rm DM} = 5 \times 10^3\,M_{\odot}$). The gravitational softenings are fixed in physical units to $\epsilon_\mathrm{DM}=42$\,pc for the dark matter and $\epsilon_\mathrm{star}=2.1$\,pc for stars. The gravitational softenings are adaptive for gas, with a minimum of $\epsilon_\mathrm{b}=0.42$\,pc. This is, again, with the exception of z5m11a and z5m11b (see Figure~\ref{fig:halos} for a list of simulation IDs considered in this work), which have $\epsilon_\mathrm{DM}=21$\,pc, $\epsilon_\mathrm{star}=1.4$\,pc, and $\epsilon_\mathrm{b}=0.28$\,pc. 

Part of the \textit{High-Redshift} suite of simulations was presented and analyzed in detail by \citet{Ma_2018_size,Ma_2018_lf} for the predicted properties of the simulated galaxy population at $5 \leq z \leq 12$, including sizes, morphologies, scaling relations, and number statistics measured by the stellar mass and luminosity functions. In this follow-up analysis of \citet{Ma_2018_lf} motivated by recent \textit{JWST} observations of the abundance of galaxies at $z\gtrsim10$, we follow closely the methodology adopted in \citet{Ma_2018_lf} for fair comparisons, but the sample size of high-$z$, massive galaxies has been substantially increased to better determine the bright-end behavior of galaxy UVLFs at cosmic dawn. Below, we will only briefly summarize the key information about the sample of simulated galaxies pertinent to the analysis presented here. We refer interested readers to the aforementioned papers for further details about the FIRE-2 simulations and the \textit{High-Redshift} suite. 

For a robust analysis of UVLFs at their bright end, we build a maximum possible sample size of massive galaxies by making use of all the zoom-in simulations available at each redshift above the ending redshift $z_\mathrm{end}$. In each zoom-in region, we consider all the well-resolved halos\footnote{Halos containing at least $10^4$ particles in total and uncontaminated by low-resolution particles are considered ``well-resolved''.} that host a \textit{central} galaxy, rather than the one hosting just the most massive, primary galaxy (typically near the center of the zoom-in region). Following \citet{Ma_2018_lf}, we define galaxies based on catalogs of halos identified with the \textsc{Amiga} Halo Finder (AHF; \citealt{KnollmannKnebe_2009}). The radius $R_\mathrm{max}$ at which the halo rotation curve reaches maximum is used to define a galaxy by incorporating star particles within $R_\mathrm{max}/3$ and excluding the contamination from subhalos outside $R_\mathrm{max}/5$. We restrict the scope of our UVLF analysis to halos with mass $M_\mathrm{h} > 10^{7.5}\,M_{\odot}$ in snapshots at $8 \leq z \leq 12$ because most of the recent UVLF measurements at $z<8$ with \textit{JWST} can be well explained by previous theoretical predictions and a sufficiently constraining sample of spectroscopically-confirmed galaxies is not available at $z>12$ \citep{Harikane_2023arXiv}. In Appendix~\ref{sec:halos}, we illustrate how the halo/galaxy sample is constructed with (snapshots of) the 26 individual zoom-in simulations, which build up a total sample of $\approx 25,000$ galaxy snapshots over $8 \leq z \leq 12$. For all simulations, snapshots are saved at a cadence of every 10--20\,Myr. 

\subsection{Processing of the Simulations} \label{sec:processing}

We process the simulated galaxy sample in order to arrive at their 1600\,\AA\ UV magnitudes $M_\mathrm{UV}$, following \citet{Sun_2023}. Templates of binary, single-stellar-population (SSP) spectra from BPASS v2.1 \citep{Eldridge_2017} are interpolated and applied to star particles according to their stellar age and metallicity, assuming a Kroupa IMF \citep{Kroupa_2001}. Including nebular (continuum) emission can in principle augment both the UV emissivity and variability \citep{Byler_2017}, although we opt to ignore it here as nebular emission is not expected to strongly affect the measurement of $M_\mathrm{UV}$, especially when compared with effects of SFR variations. Two notable differences from \citet{Sun_2023} exist, though, for the treatment of (1) the connection between $M_\mathrm{UV}$ and the SFH and (2) dust attenuation, on which we elaborate below. 

\subsubsection{Bursty vs Smoothed Star Formation Histories} \label{sec:def_bursty_vs_smoothed}

At cosmic dawn, an increased SFR variability can strongly modulate the observed number statistics of galaxies. To assess the impact of bursty star formation on the $M_\mathrm{UV}$--$M_\mathrm{h}$ relation and thus the UVLF, we consider two contrasting scenarios to model $M_\mathrm{UV}$. 

The baseline scenario, which we refer to as ``bursty'', assumes that the SFH of each galaxy in our sample is exactly as predicted by the simulations and thus $M_\mathrm{UV}$ can be derived by summing up the spectral emissivities of all star particles of the galaxy at a given snapshot according to their age and metallicity, as in \citet{Sun_2023}. This is the approach most faithful to the SFHs predicted by the simulations. In this approach, $M_\mathrm{UV}$ naturally inherits the burstiness predicted by the simulations --- as the SFR varies, the UV 1600\,\AA\ luminosity of the galaxy also fluctuates accordingly because most of FUV continuum emission is sourced by the massive, short-lived stars formed. As a result, a bursty SFH imprints significant stochasticity in $M_\mathrm{UV}$ at a fixed stellar or halo mass. 

In the contrasting scenario, which we refer to as ``smoothed'', we artificially reduce the impact of bursty SFH on the evaluation of $M_\mathrm{UV}$ by redistributing the ages of star particles (while retaining their metallicities). Specifically, we first define a smoothing kernel of duration $\tau_\mathrm{SF}$ Myr and bin star particles using their star formation times into time bins of width $\tau_\mathrm{SF}$. We then redistribute the ages of the star particles in individual bins such that the stellar mass forms at a nearly constant rate by enforcing evenly-distributed star formation times within each bin. This redistribution of stellar ages effectively smooths the SFH and reduces to the ``bursty'' case for a sufficiently small $\tau_\mathrm{SF}$. Notably, unlike some previous work where effects of varying the UV variability on UVLFs are studied assuming a fixed mean/median $L_\mathrm{UV}$--$M_\mathrm{h}$ relation \cite[e.g.,][]{MF_2023,Shen_2023}, our method by its nature conserves the total amount of cosmic star formation such that the two scenarios differ only in terms of the short-timescale SFR variability and its impact on the UV emissivity. 

\subsubsection{Dust Attenuation} \label{sec:dust}

Observations have shown compelling evidence of early chemical enrichment and the production of non-negligible dust in galaxies at $z \gtrsim 7$ \citep{Tamura_2019,Fudamoto_2021,Witstok_2023}. A reasonable treatment of dust attenuation is therefore needed for our predictions of the UVLF at cosmic dawn, especially at the bright end because massive (intrinsically UV-bright) galaxies generally contain more dust. 

To estimate the effect of dust attenuation on $M_\mathrm{UV}$, we employ an empirical model motivated by an up-to-date measurement of the $\beta_\mathrm{UV}$--$M_\mathrm{UV}$ (color--magnitude) relation at $z>8$ by \cite{Cullen_2023} using a combination of \textit{JWST} and ground-based observations\footnote{See also \citet{Topping_2023}, who find slightly steeper $\beta_\mathrm{UV}$ that steepens with increasing redshift from $z\sim6$--12.}. We combine the best-fit relation $\beta_\mathrm{UV} = -0.17 M_\mathrm{UV} + 5.40$ with the attenuation--UV slope relation, $A_\mathrm{UV} = 0.48 (\beta_\mathrm{UV} + 2.62)$, determined from $z\approx5.5$ galaxies observed in the ALPINE survey (\citealt{Fudamoto_2020}; see also \citealt{Reddy_2018}). While an extrapolation in redshift is involved, this best-fit relation from ALMA observations represents a state-of-the-art empirical baseline for estimating dust attenuation properties at cosmic dawn, which should suffice for the purpose of this work. We neglect the scatter around these mean relations given its small impact on $M_\mathrm{UV}$ and caution that results with dust attenuation included that follow should be taken as rough estimates only. The validity of these simplistic treatments can be tested with simulations with detailed dust radiative transfer \citep{Cochrane_2019,Cochrane_2022,Ma_2019,Vogelsberger_2020,Shen_2022} and multi-wavelength observations \citep{Akins_2023, Bakx_2023}, which are left for future work. We note, though, that at $z>10$ the difference between UVLFs with and without dust attenuation is predicted to be very small in our model (see Figure~\ref{fig:uvlf_comparison}), such that uncertainties in the treatment of dust should not affect our results significantly. 

\subsection{Estimating the UVLF from Zoom-in Simulations} \label{sec:estimate_uvlf}

Using UV magnitudes derived for the sample of simulated galaxies binned into redshift bins of width $\Delta z = \pm 0.5$, we calculate the UVLF through a convolution with the halo mass function (HMF) following the ``HMF-weighting'' method introduced by \citet{Ma_2018_lf}. This method has been verified to provide robust estimates of the UVLF from galaxy samples drawn from zoom-in simulations, so we only summarize briefly here. First, in narrow halo mass and redshift bins, we count the number of simulated halos $N_\mathrm{S}$ from the sample and compute the expected number of halos $N_\mathrm{E}$, which scales with the HMF, $\mathrm d n / \mathrm d \log M_\mathrm{h}$, calculated using the \textsc{hmf} code \citep{Murray_2013} for the fitting function from \citet{Behroozi_2013}. A common weight $w = N_\mathrm{E}/N_\mathrm{S}$ is assigned to all the halos in the same bin, such that a summation of halo weights in a given mass bin yields the expected number of halos in the universe. These weights are then applied to sample galaxies binned in $M_\mathrm{UV}$ to obtain the UVLF, which is essentially a convolution between the HMF and $M_\mathrm{UV}$--$M_\mathrm{h}$ relation including the full, $M_\mathrm{h}$-dependent distribution \cite[see Section 2.4 of][]{Ma_2018_lf}. Finally, we stress that, compared with \citet{Ma_2018_lf} where only a subset of the \textit{High-Redshift} suite was analyzed, we substantially increase the number of samples of massive halos/bright galaxies in this work (a factor 8 increase of halos with $M_\mathrm{h}>10^{10}\,M_{\odot}$ at $z=10$) by considering the full \textit{High-Redshift} suite as in \citet{Ma_2019}, thereby extending the magnitude down to which the UVLF at $z>10$ can be reliably determined to $M_\mathrm{UV} < -20$, overlapping with the bright-end UVLF probed by \textit{JWST}. 

\section{Results} \label{sec:results}

\begin{figure*}
    \centering
	\includegraphics[width=\textwidth]{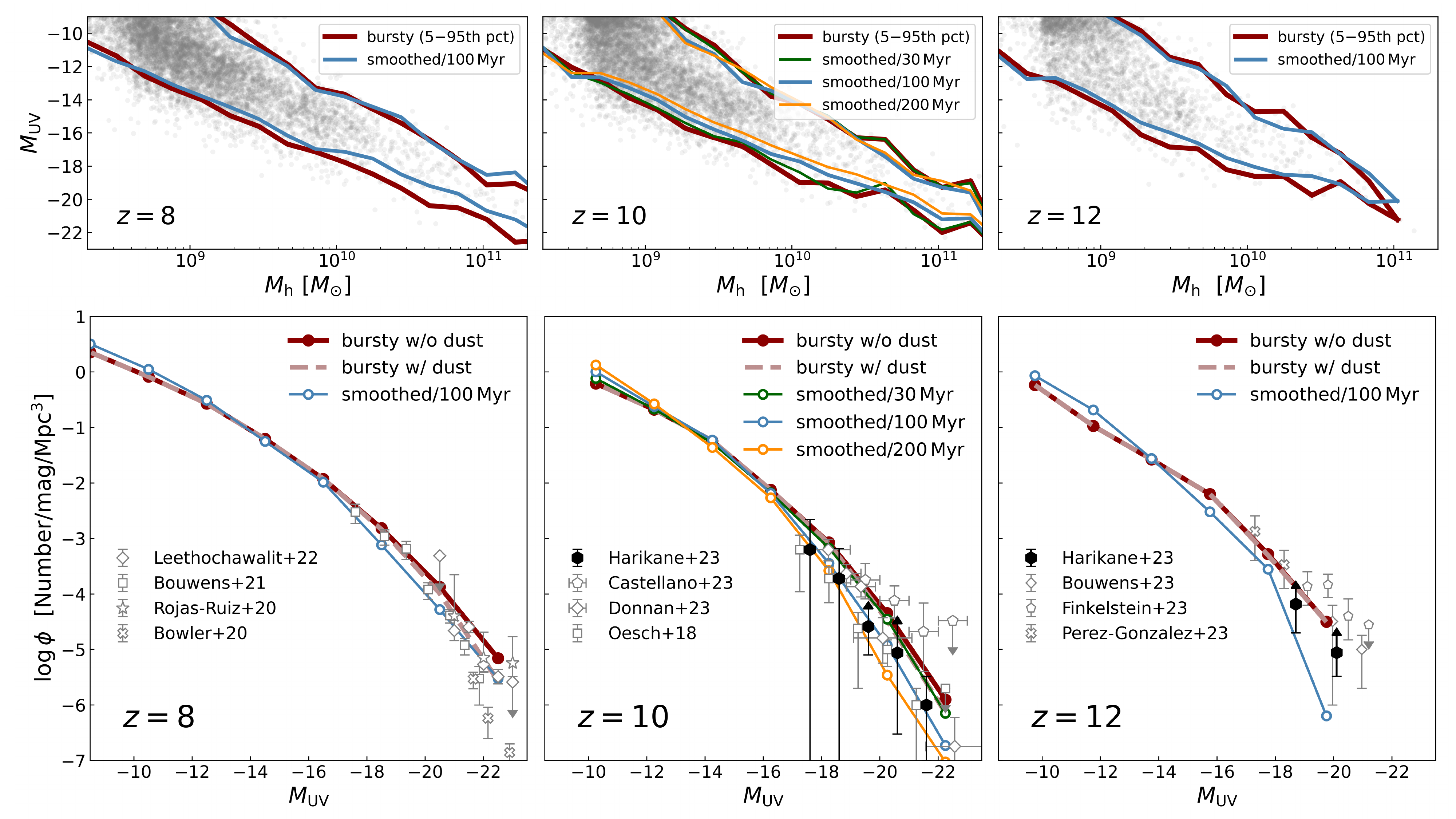}
    \caption{\textit{Top:} UV magnitude--halo mass relations at $z=8$--12. Data for individual galaxies are denoted by the grey dots (no smoothing applied to the SFH). The thick solid curves indicate the range of the 5th and 95th percentiles in the ``bursty'' and ``smoothed'' cases, from which the suppression of bright galaxy number counts due to smoothing is apparent. \textit{Bottom:} UVLFs at $z=8$--12 derived from the convolution between the UV magnitude--halo mass relation and the HMF. Dust-free predictions are shown as solid for both ``bursty'' and ``smoothed'' cases, whereas the dust-attenuated scenario is shown as dashed for only the ``bursty'' case (Section~\ref{sec:dust}) for visual clarity. Constraints from observations are shown by the data points in black for the spectroscopically-confirmed-only samples \citep{Harikane_2023arXiv} and in grey for data sets involving photometric candidates \citep{Oesch_2018, Bowler_2020, Rojas-Ruiz_2020, Bouwens_2021, Bouwens_2023, Finkelstein_2022, Leethochawalit_2022, Castellano_2023, Donnan_2023, Harikane_2023arXiv, Perez-Gonzalez_2023}. Cases with larger and smaller smoothing timescale $\tau_\mathrm{SF}$ values than the fiducial one (100\,Myr) are shown at $z=10$ to illustrate the impact of SFH smoothing on the UVLF.}
    \label{fig:master}
\end{figure*}

\subsection{The $M_\mathrm{UV}$--$M_\mathrm{h}$ Relation and the UVLF}

Following the methods outlined in Sections~\ref{sec:processing} and \ref{sec:estimate_uvlf}, we first use our samples of simulated galaxies to quantify the $M_\mathrm{UV}$--$M_\mathrm{h}$ relation in different redshift regimes, assuming either ``bursty'' or ``smoothed'' SFH. A comparison of the $M_\mathrm{UV}$--$M_\mathrm{h}$ relations at $z=8$, 10, and 12 from our simulations is shown in the top row of Figure~\ref{fig:master}. Overall, galaxies become more UV-bright at higher $M_\mathrm{h}$ and, at a given $M_\mathrm{h}$, $M_\mathrm{UV}$ decreases modestly with increasing redshift as a result of more rapid halo growth at higher redshift. A significant scatter in $M_\mathrm{UV}$ around the median relation that gradually increases towards lower masses exists, which is a sign of increasing star formation burstiness at low masses, given the proportionality between $L_\mathrm{UV}$ and the SFR. At a fixed $M_\mathrm{h}$, we find a modest trend for the scatter in $M_\mathrm{UV}$ to decrease with decreasing redshift that continues to $z<8$ (not shown). This tentative evidence for the redshift evolution of the UV variability might be testable using comparisons of SFR indicators sensitive to different star formation timescales or    high-precision measurements of the halo--galaxy connection with galaxy clustering (see Section~\ref{sec:discussion}). From the comparison between the ``bursty'' and ``smoothed'' cases shown by the 5--95th percentiles (especially in the top middle panel where three ``smoothed'' cases with varying $\tau_\mathrm{SF}$ are shown), it can be seen that evaluating $M_\mathrm{UV}$ from a smoothed SFH leads to a shallower $M_\mathrm{UV}$--$M_\mathrm{h}$ relation with a reduced scatter in $M_\mathrm{UV}$ at higher masses, which effectively suppresses the population of UV-bright galaxies at a given $M_\mathrm{h}$. 

In the bottom row of Figure~\ref{fig:master}, we show the UVLF at $z=8$--12 implied by the $M_\mathrm{UV}$--$M_\mathrm{h}$ relation. From the comparisons against recent observational constraints and between the two SFH cases, several key results are immediately apparent. First, in the fiducial, ``bursty'' SFH scenario, the predicted UVLFs agree remarkably well with the observational constraints available. In particular, our $z \gtrsim 10$ predictions lie safely above the firm lower bounds set by the dust-uncorrected, spectroscopically-confirmed samples recently compiled by \citet{Harikane_2023arXiv}, and they are also broadly consistent with the variety of measurements based on photometrically selected candidates (see the caption for details)\footnote{We have verified by bootstrapping 1000 times the simulated galaxy samples that the statistical uncertainty on the UVLF, especially at the bright end, is small enough that it does not affect the bright-end comparisons of interest to this study. In the brightest bin, the 1$\sigma$ statistical uncertainties in $\log \phi$ estimated from bootstrapping are approximately 0.15\,dex, 0.15\,dex, and 0.3\,dex at $z=8$, 10, and 12, respectively.}. Unlike some other theoretical predictions \cite[e.g.,][]{Mason_2023, Yung_2023}, for which a clear tension with the spec-$z$ lower bounds exists without modifications, our bursty-case predictions do not require any additional tuning of UV variability or production efficiency to match observations. Despite uncertainties associated with the treatment of dust, this good agreement implies that the UVLFs observed by \textit{JWST} at $z \gtrsim 10$ are consistent with generally ``normal'' SFE and UV production efficiency as predicted by the FIRE-2 simulations. As demonstrated in \citet{Ma_2018_lf}, the relation between $M_*$ and $M_\mathrm{h}$ in these simulations is broadly consistent with extrapolations from lower $z$ where empirical analyses show that the SFE is strongly suppressed by stellar feedback in low-mass halos \citep{Behroozi_2013, Tacchella_2018}. 

\begin{table}
\centering
\caption{Dust-free UVLFs at $z=8$, 10, and 12 from the simulated galaxies.}
\begin{tabular}{cccccc}
\hline
$M_\mathrm{UV}$ & $\log \phi$ & $M_\mathrm{UV}$ & $\log \phi$ & $M_\mathrm{UV}$ & $\log \phi$ \\
\hline
\multicolumn{2}{c}{$z=8$} & \multicolumn{2}{c}{$z=10$} & \multicolumn{2}{c}{$z=12$} \\
\cmidrule(l{2em}r{2em}){1-2}
\cmidrule(l{2em}r{2em}){3-4} 
\cmidrule(l{2em}r{2em}){5-6}
$-10.5$ & $-0.085$ & $-10.25$ & $-0.207$ & $-9.75$ & $-0.234$ \\
$-12.5$ & $-0.570$ & $-12.25$ & $-0.677$ & $-11.75$ & $-0.971$ \\
$-14.5$ & $-1.206$ & $-14.25$ & $-1.242$ & $-13.75$ & $-1.576$ \\
$-16.5$ & $-1.926$ & $-16.25$ & $-2.124$ & $-15.75$ & $-2.200$ \\
$-18.5$ & $-2.815$ & $-18.25$ & $-3.072$ & $-17.75$ & $-3.282$ \\
$-20.5$ & $-3.872$ & $-20.25$ & $-4.344$ & $-19.75$ & $-4.500$ \\
$-22.5$ & $-5.158$ & $-22.25$ & $-5.902$ & $$ & $$ \\
\hline
\end{tabular}
\begin{tablenotes}
\item \textbf{Notes.} \\ $\phi$ values are quoted in units of $\mathrm{mag^{-1}\,Mpc^{-3}}$. See Equation~(\ref{eq:dpl}) for analytic fits to the UVLF over $8 < z < 12$. For reference, in the two brightest bins, $\phi$ is extracted from a sample of (39, 17), (39, 13), (93, 17) galaxies at $z=8$, 10, and 12, respectively. 
\end{tablenotes}
\label{tb:uvlf}
\end{table}

Second, in the contrasting, ``smoothed'' SFH scenario, a clear deficit of UV-bright galaxies is seen as a result of suppressed up-scattering in $M_\mathrm{UV}$ of low-mass halos when the SFR is averaged over a long timescale $\tau_\mathrm{SF}$. The underestimated abundance of UV-bright galaxies reveals the important role played by the burstiness of star formation in determining the number statistics of galaxies at cosmic dawn. As also shown by the comparison of different $\tau_\mathrm{SF}$ values at $z=10$, smoothed SFHs with $\tau_\mathrm{SF} \gtrsim 100\,$Myr result in bright-end UVLFs that are too steep compared with observations, especially the photometrically selected samples, for which the bright-end UVLF can be underpredicted at $>2\sigma$ level in some cases \citep{Castellano_2023, Donnan_2023}. At $z=12$, predictions of the smoothed SFH are in tension with even the most conservative lower limits derived from only the spectroscopically confirmed samples \citep{Harikane_2023arXiv}. It is therefore clear that the UVLF serves as a useful probe of the burstiness in the SFH, as been noted in e.g., \citet{FM_2022} and \citet{Shen_2023}, although in practice it can be challenging to extract the burstiness information from only the UVLF measurements (see Section~\ref{sec:discussion}). The overall shallower $M_\mathrm{UV}$--$M_\mathrm{h}$ relation when the SFH is smoothed also leads to slightly steeper slope at the faint end, although the effect is much smaller than the suppression at the bright end. 

\begin{figure}
    \centering
	\includegraphics[width=\columnwidth]{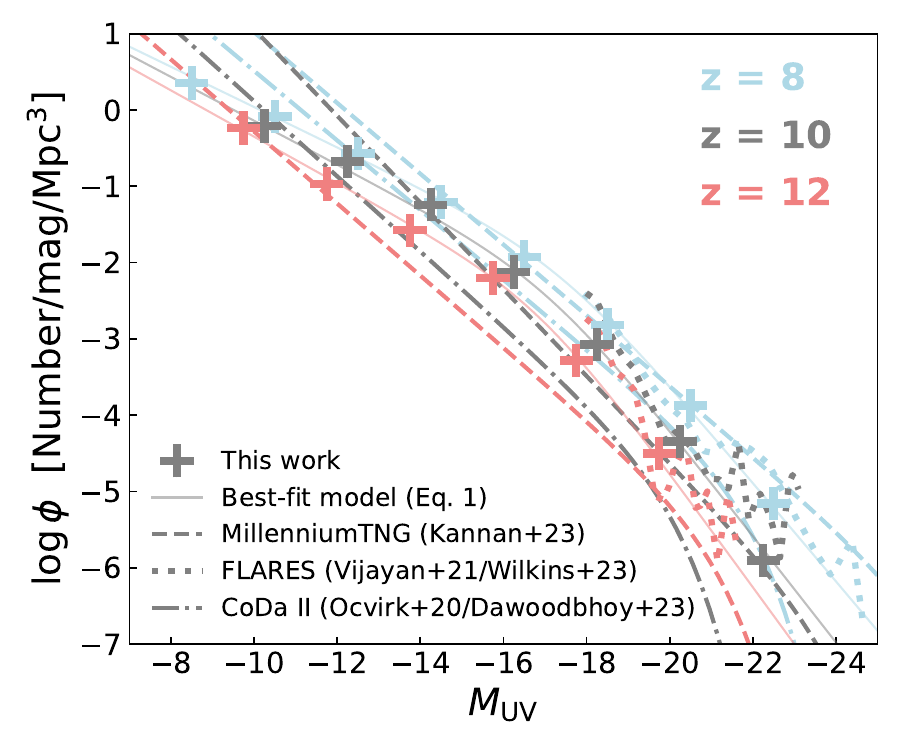}
    \caption{Dust-free UVLFs at $z=8$, 10, and 12 predicted by the FIRE-2 simulations and from the literature. The binned and the best-fit, double-power law UVLFs are denoted by the crosses and solid curves, as specified in Table~\ref{tb:uvlf} and Equation~(\ref{eq:dpl}), respectively. Several example dust-free predictions from other cosmological hydrodynamical simulations, including MillenniumTNG \citep[dashed,][]{Kannan_2023}, FLARES \citep[dotted,][]{Vijayan_2021, Wilkins_2023}, and CoDa~II \citep[dotted and only at $z=8$ and 10,][]{Ocvirk_2020,Dawoodbhoy_2023} are also plotted for comparison.}
    \label{fig:uvlf_comparison}
\end{figure}

The binned UVLFs without dust attenuation extracted from our simulations at $z=8$, 10, and 12 are summarized in Table~\ref{tb:uvlf}. As has been demonstrated in Figure~\ref{fig:master}, dust attenuation only modestly affects the UVLF at the very bright end, reducing $\phi$ (in the brightest bin) by approximately 0.4, 0.25, and 0.01\,dex at $z=8$, 10, and 12, respectively. The binning scheme is chosen such that the brightest $M_\mathrm{UV}$ bin contains more than ten simulated galaxies for robust statistics. Meanwhile, we fit the dust-free UVLF at $8 \leq z \leq 12$ assuming a universal double-power law (DPL) in $M_\mathrm{UV}$, 
\begin{equation}
\Phi(M_\mathrm{UV}) = \frac{0.4 (\ln10)\, 10^{\phi_{*}}}{10^{0.4(\alpha+1)(M^{*}_\mathrm{UV}-M_\mathrm{UV})} + 10^{0.4(\beta+1)(M^{*}_\mathrm{UV}-M_\mathrm{UV})}}.
\label{eq:dpl}
\end{equation}
We specify the redshift-dependent DPL parameters $\phi_{*}$, $M^{*}_\mathrm{UV}$, $\alpha$, and $\beta$ in the form of a single power law as $\phi_{*}(z) = \phi_{*,0}[(1+z)/10]^{\phi_{*,1}}$, $M^{*}_\mathrm{UV}(z) = M^{*,0}_\mathrm{UV}[(1+z)/10]^{M^{*,1}_\mathrm{UV}}$, $\alpha_{*}(z) = \alpha_{*,0}[(1+z)/10]^{\alpha_{*,1}}$, and $\beta_{*}(z) = \beta_{*,0}[(1+z)/10]^{\beta_{*,1}}$, where the best-fit parameters are found to be $\phi_{*,0} = -2.01$, $\phi_{*,1} = 0.68$, $M^{*,0}_\mathrm{UV} = -17.26$, $M^{*,1}_\mathrm{UV} = -0.08$, $\alpha_{*,0} = -0.31$, $\alpha_{*,1} = -0.93$, $\beta_{*,0} = 0.68$, and $\alpha_{*,1} = 0.93$. 

Figure~\ref{fig:uvlf_comparison} shows a comparison between the binned and best-fit UVLFs predicted by our simulations and other theoretical predictions in the literature based on cosmological hydrodynamical simulations \citep{Ocvirk_2020, Vijayan_2021, Dawoodbhoy_2023, Kannan_2023, Wilkins_2023}. Overall, our predicted UVLFs show a weaker redshift evolution beyond $z = 8$ compared with the predictions from the MillenniumTNG \citep{Kannan_2023} and CoDa~II \citep{Ocvirk_2020, Dawoodbhoy_2023} simulations, which results in a higher abundance of bright ($M_\mathrm{UV} \lesssim -20$) galaxies at $z \gtrsim 10$. Our bright-end predictions are generally comparable to those from the FLARES simulations \citep{Vijayan_2021, Wilkins_2023} in both normalization and slope, despite the vastly different nature of the simulations and methods to evaluate the UVLF. It is noteworthy, though, that the FIRE-2 simulations analyzed in this work have significantly higher resolution ($m_\mathrm{b} \approx 7 \times 10^3\,M_{\odot}$ in FIRE-2 vs. $m_\mathrm{b} \approx 2 \times 10^6\,M_{\odot}$ in FLARES), which allows us to predict the UVLFs at $8 \leq z \leq 12$ down to $M_\mathrm{UV} \sim -10$ vs. the FLARES predictions down to $M_\mathrm{UV} \sim -18$. We have also verified that UVLFs in this work and from \citet{Ma_2018_lf, Ma_2019} are in good agreement in the overlapping regime. 

\subsection{UV Luminosity Density}

\begin{figure}
    \centering
	\includegraphics[width=\columnwidth]{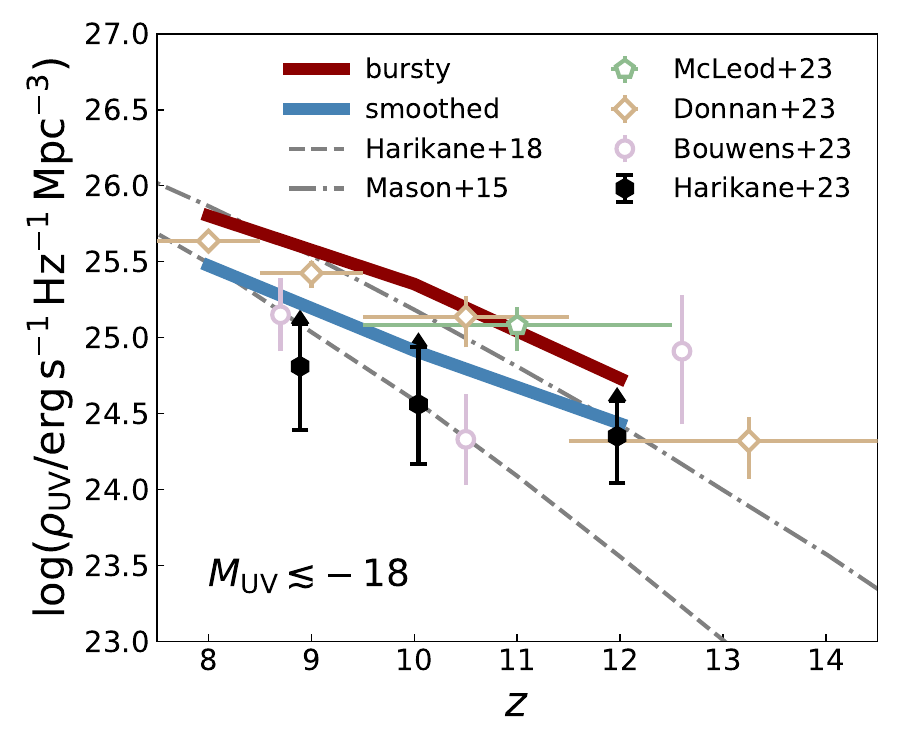}
    \caption{The cumulative UV luminosity density $\rho_\mathrm{UV}(<M_\mathrm{UV,lim})$ integrated down to $M_\mathrm{UV,lim} \simeq -18$ with dust attenuation included (see Section~\ref{sec:dust}). At $z\gtrsim10$, some theoretical models \citep[e.g.,][]{Mason_2015, Harikane_2018} underestimate $\rho_\mathrm{UV}$ compared with observational constraints based on photometric galaxy candidates \citep[e.g.,][]{Bouwens_2023, Donnan_2023, McLeod_2023, Perez-Gonzalez_2023} and/or spectroscopically-confirmed galaxies as firm lower limits \citep{Harikane_2023arXiv}. Predictions from our ``bursty'' case are broadly consistent with both photometric and spectroscopic samples and show a slightly weaker redshift evolution $\rho_\mathrm{UV} \propto (1+z)^{-0.3}$ over $8 \leq z \leq 12$.}
    \label{fig:rhouv}
\end{figure}

By integrating the predicted UVLFs, we can derive the UV luminosity density, $\rho_\mathrm{UV}$, as a function of time, which traces the cosmic star formation rate density (SFRD). Since at $z \gtrsim 10$ only the brightest end ($M_\mathrm{UV} \ll M_\mathrm{UV,*}$) of the UVLF has been probed, we follow \citet{Harikane_2023arXiv} to compare the UV luminosity density contributed by galaxies brighter than $M_\mathrm{UV} = -18$, namely $\rho_\mathrm{UV,bright} = \rho_\mathrm{UV}(M_\mathrm{UV} < -18)$, which corresponds to the contribution from halos with $M_\mathrm{h} \gtrsim 10^{10}\,M_{\odot}$ at $z=10$. The unconstrained contribution by fainter, lower-mass galaxies is highly sensitive to the faint-end slope of the UVLF and might even outweigh $\rho_\mathrm{UV,bright}$ \citep{SF_2016}, but the comparison restricted to $M_\mathrm{UV} < -18$ galaxies still serves as a useful test of the overall abundance of bright, massive galaxies and their SFE at cosmic dawn\footnote{Results from this work, \citet{Harikane_2018, Harikane_2023ApJS}, and \citet{Bouwens_2023} are integrated down to $M_\mathrm{UV,lim} = -18$, whereas the rest are down to $M_\mathrm{UV,lim} = -17$. Figure~\ref{fig:rhouv} thus shows conservatively that our simulations without smoothing predict enough total UV emission compared with observations, regardless of the modest difference in $M_\mathrm{UV,lim}$.}. 

Figure~\ref{fig:rhouv} shows a comparison of the cumulative UV luminosity density between the dust-attenuated predictions from our simulations and a compilation of constraints from observations and theoretical forecasts in the literature. Throughout, dust-attenuated predictions from models/simulations (curves) are compared with observations (data points), which are dust-uncorrected. Over $8 \leq z \leq 12$, dust-attenuated luminosity densities predicted by our simulations without smoothing the SFH are fully consistent with observations of both photometric galaxy candidates and spectroscopically-confirmed galaxies that provide firm lower limits. Due to the integrated nature of $\rho_\mathrm{UV}$, the ``smoothed'' case appears more consistent with the spec-$z$-only lower limits here than at the bright end of the UVLF as shown in Figure~\ref{fig:uvlf_comparison}. In both cases with dust attenuation, a power-law evolution of $\rho_\mathrm{UV} \propto (1+z)^{-0.3}$ over $8 \leq z \leq 12$ is implied, which appears more gradual compared with the predictions by some previously proposed semi-analytic/semi-empirical models, such as in \citet{Mason_2015} and \citet{Harikane_2018}. 

\section{Discussion and Conclusions} \label{sec:discussion}

We have demonstrated that the FIRE-2 simulations with a multi-channel implementation of standard stellar feedback processes can reproduce well the observed abundance of UV-bright galaxies at $z\gtrsim10$, including both the photometrically selected candidates and the spectroscopically confirmed sources recently discovered by \textit{JWST}. We further showed that the bursty SFH predicted to be common in galaxies at cosmic dawn is important for explaining the bright-end of the UVLF. 
With burstiness included, the simulations demonstrate that a boosted UV emissivity due to, e.g., an enhanced SFE, a top-heavy IMF, AGN contributions, or Population~III stars \cite[see e.g.,][]{Harikane_2023ApJS, Harikane_2023AGN}, is not necessary to explain the bright-end UVLF at $z\gtrsim10$. 
(This is of course not to say that none of these other effects could be present in the real universe, so it certainly remains interesting to investigate these other possibilities!) Compared to semi-analytic/empirical models \cite[][]{Mason_2023,MF_2023,Shen_2023,Yung_2023}, our predictions based on the FIRE-2 simulations avoid ad hoc fine-tuning of the $M_\mathrm{UV}$--$M_\mathrm{h}$ relation to match observations. 

Though not shown explicitly in this Letter, we have verified that the stellar mass--halo mass (SMHM) relation, as a measure of the time-integrated, galaxy-scale SFE, $f_{\star} \equiv M_{\star}/(f_\mathrm{b} M_\mathrm{h}$) (where $f_\mathrm{b}=\Omega_{\rm b}/\Omega_{\rm m}$ is the cosmic baryon fraction), barely evolves over $5 \leq z \leq 12$ in our simulations. This is consistent with the previous results presented in \citet{Ma_2018_lf} based on a subset of the full \textit{High-Redshift} suite considered in this work (see their Figure~4). 
These results indicate that $f_{\star}$ changes from approximately $10^{-3.3}$ to $10^{-1.5}$ as $M_\mathrm{h}$ increases from $10^{8}\,M_{\odot}$ to $10^{11}\,M_{\odot}$ following a simple power law of slope $\sim0.6$ in log-log space. 
Thus, even though star formation is bursty, the galaxy-scale SFE is not strongly enhanced in these simulations relative to, e.g., an extrapolation of the SMHM relation empirically determined at lower redshift \citep[][]{Behroozi_2019}. 
In particular, our simulations do not appear to realize the ``feedback-free starburst'' scenario predicted by \cite{Dekel_2023} using analytic arguments, which would result in $f_{\star}$ values up to order-unity.\footnote{While the FIRE-2 simulations assume a local, \textit{instantaneous} SFE of 100\% per free-fall time, this only applies in dense, self-gravitating gas \citep[see the methods in][]{Hopkins_2018}. 
On galaxy and molecular cloud scales, stellar feedback generally regulates the SFE to much lower values \citep[e.g.,][]{Grudic_2018, Orr_2018, Gurvich_2020}.}

We note that \citet{PallottiniFerrara_2023} also recently used a set of cosmological zoom-in simulations (SERRA; \citealt{Pallottini_2022}) to investigate some implications of stochastic star formation in early galaxies for the abundance of $z\gtrsim10$ galaxies observed by \textit{JWST}. By characterizing the distribution of time-dependent variations in the SFR of individual galaxies, they concluded that the predicted SFR variability cannot account for the required boost suggested by some recent literature to match the observed UVLF at $z\gtrsim10$ \citep{MF_2023, Shen_2023}. However, \citet{PallottiniFerrara_2023} did not self-consistently derive the UVLF from their simulations. Since other physical factors such as the SFE also impact the UVLF, in addition to burstiness \citep{MF_2023, Munoz_2023}, in order to unambiguously gauge the importance of bursty star formation it is desirable to perform a self-consistent, end-to-end study of the UVLF as we do in this work.

Looking ahead, a detailed characterization of the SFR variability on different timescales will shed light on the physical processes at play in the build-up of galaxies at early times, as has been demonstrated in recent work using periodogram \citep{PallottiniFerrara_2023} or more generally power spectral density (PSD) analysis \citep{Iyer_2020,Tacchella_2020}. Moreover, various implications of bursty star formation should be explicitly considered when interpreting observations of high$-z$ galaxies. 
For example, \cite{Sun_2023} showed that SFR variability introduces important selection effects in rest UV-selected samples. 
Since most galaxies at cosmic dawn may form stars in a highly bursty manner, the impact of burstiness on galaxy number statistics also raises questions about how to reliably constrain cosmology with high-$z$ galaxy observations \citep{Sabti_2023}. 
At the same time, it is of great interest to investigate how to observationally characterize the time variability of star formation and its mass and redshift dependence, e.g. using SFR indicators sensitive to different timescales \citep{Sparre_2017, Flores_2021, Sun_2023_EBL} or the spatial clustering of galaxies \citep{Munoz_2023}. 
Quantifying the effects of bursty star formation on statistics such as galaxy clustering is a critical stepping stone towards the usage of high-$z$ galaxies as robust cosmological probes. \\

The authors thank the anonymous reviewer for comments that helped improve this Letter, as well as Pratik Gandhi, Yuichi Harikane, and Julian Mu\~{n}oz for helpful discussion. GS was supported by a CIERA Postdoctoral Fellowship. CAFG was supported by NSF through grants AST-2108230  and CAREER award AST-1652522; by NASA through grants 17-ATP17-0067 and 21-ATP21-0036; by STScI through grant HST-GO-16730.016-A; and by CXO through grant TM2-23005X. The Flatiron Institute is supported by the Simons Foundation. AW received support from: NSF via CAREER award AST-2045928 and grant AST-2107772; NASA ATP grant 80NSSC20K0513; and HST grants AR-15809, GO-15902, GO-16273 from STScI. 
The simulations used in this Letter were run on XSEDE computational resources (allocations TG-AST120025, TG-AST130039, TG-AST140023, and TG-AST140064). Additional analysis was done using the Quest computing cluster at Northwestern University.

\software{\textsc{BPASS} \citep{Eldridge_2017}, \textsc{GizmoAnalysis} \citep{Wetzel_2016,2020ascl.soft02015W}, \textsc{hmf} \citep{Murray_2013}}

\appendix

\twocolumngrid

\section{Forming the halo/galaxy sample} \label{sec:halos}

\begin{figure}
    \centering
	\includegraphics[width=\columnwidth]{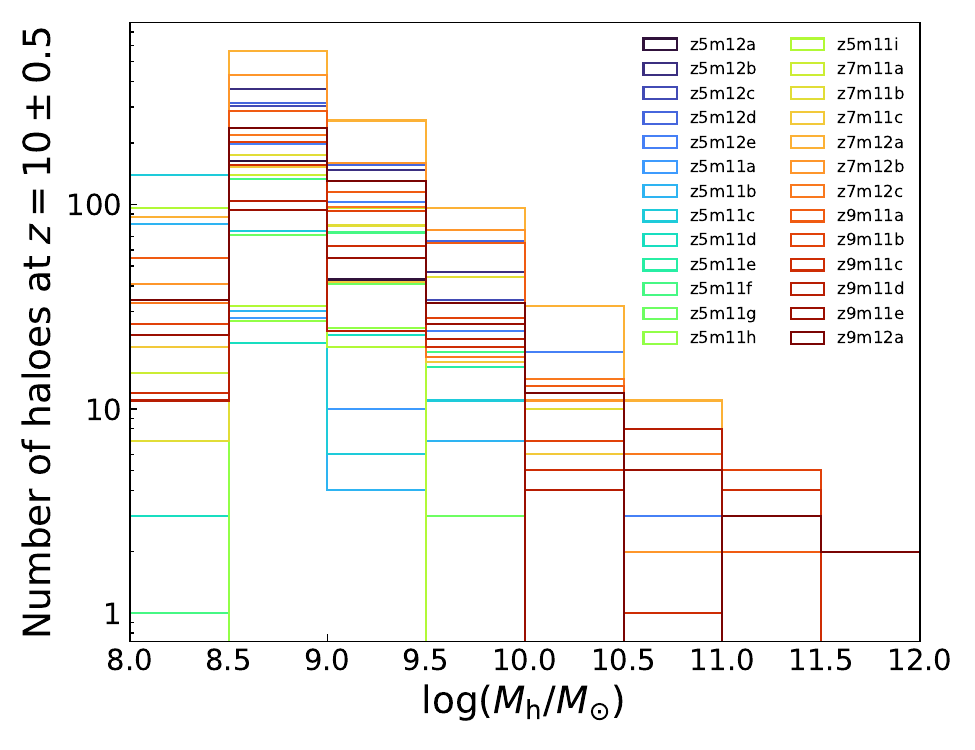}
    \caption{The number of galaxies sampled in each halo mass bin for the 26 simulations inspected for evaluating the UVLF at $z=10 \pm 0.5$. This amounts to a total of $\approx 9000$ galaxies sampled from snapshots of the 26 zoom-in regions over $9.5 < z < 10.5$. Simulation IDs are listed in the legend \citep[c.f.,][]{Sun_2023}.}
    \label{fig:halos}
\end{figure}

Throughout, we analyze snapshots of a galaxy in a $\Delta z = 0.5$ bin multiple times per the cadence at which snapshots are stored (every 10--20\,Myr). While the same galaxy from neighbouring snapshots are not strictly independent as far as $M_\mathrm{UV}$ is considered, this method is useful because the highly time-variable SFR limits the temporal correlation between consecutive snapshots. It yields a large statistical sample appropriate for UVLF analysis (see Figure~\ref{fig:halos}) and the sampling cadence does not bias the results, as have been shown by analyses that randomly exclude approximately half of the samples \citep{Ma_2018_lf}. At $z=8$, 10, and 12, the UVLF is evaluated from a sample of approximately 12,000, 9,000, and 4,000 galaxies, respectively. Summing over the three redshift bins, this yields $\approx 25,000$ galaxy samples in total.

\bibliography{jwst_uvlf}{}

\begin{thebibliography}{}
\expandafter\ifx\csname natexlab\endcsname\relax\def\natexlab#1{#1}\fi
\providecommand{\url}[1]{\href{#1}{#1}}
\providecommand{\dodoi}[1]{doi:~\href{http://doi.org/#1}{\nolinkurl{#1}}}
\providecommand{\doeprint}[1]{\href{http://ascl.net/#1}{\nolinkurl{http://ascl.net/#1}}}
\providecommand{\doarXiv}[1]{\href{https://arxiv.org/abs/#1}{\nolinkurl{https://arxiv.org/abs/#1}}}

\bibitem[{{Akins} {et~al.}(2023){Akins}, {Casey}, {Allen}, {Bagley}, {Dickinson}, {Finkelstein}, {Franco}, {Harish}, {Arrabal Haro}, {Ilbert}, {Kartaltepe}, {Koekemoer}, {Liu}, {Long}, {McCracken}, {Paquereau}, {Papovich}, {Pirzkal}, {Rhodes}, {Robertson}, {Shuntov}, {Toft}, {Yang}, {Barro}, {Bisigello}, {Buat}, {Champagne}, {Cooper}, {Costantin}, {de la Vega}, {Drakos}, {Faisst}, {Fontana}, {Fujimoto}, {Gillman}, {G{\'o}mez-Guijarro}, {Gozaliasl}, {Hathi}, {Hayward}, {Hirschmann}, {Holwerda}, {Jin}, {Kocevski}, {Kokorev}, {Lambrides}, {Lucas}, {Magdis}, {Magnelli}, {McKinney}, {Mobasher}, {P{\'e}rez-Gonz{\'a}lez}, {Rich}, {Seill{\'e}}, {Talia}, {Urry}, {Valentino}, {Whitaker}, {Yung}, \& {Zavala}}]{Akins_2023}
{Akins}, H.~B., {Casey}, C.~M., {Allen}, N., {et~al.} 2023, arXiv e-prints, arXiv:2304.12347, \dodoi{10.48550/arXiv.2304.12347}

\bibitem[{{Arrabal Haro} {et~al.}(2023){Arrabal Haro}, {Dickinson}, {Finkelstein}, {Kartaltepe}, {Donnan}, {Burgarella}, {Carnall}, {Cullen}, {Dunlop}, {Fern{\'a}ndez}, {Fujimoto}, {Jung}, {Krips}, {Larson}, {Papovich}, {P{\'e}rez-Gonz{\'a}lez}, {Amor{\'\i}n}, {Bagley}, {Buat}, {Casey}, {Chworowsky}, {Cohen}, {Ferguson}, {Giavalisco}, {Huertas-Company}, {Hutchison}, {Kocevski}, {Koekemoer}, {Lucas}, {McLeod}, {McLure}, {Pirzkal}, {Trump}, {Weiner}, {Wilkins}, \& {Zavala}}]{ArrabalHaro_2023}
{Arrabal Haro}, P., {Dickinson}, M., {Finkelstein}, S.~L., {et~al.} 2023, arXiv e-prints, arXiv:2303.15431, \dodoi{10.48550/arXiv.2303.15431}

\bibitem[{{Bakx} {et~al.}(2023){Bakx}, {Zavala}, {Mitsuhashi}, {Treu}, {Fontana}, {Tadaki}, {Casey}, {Castellano}, {Glazebrook}, {Hagimoto}, {Ikeda}, {Jones}, {Leethochawalit}, {Mason}, {Morishita}, {Nanayakkara}, {Pentericci}, {Roberts-Borsani}, {Santini}, {Serjeant}, {Tamura}, {Trenti}, \& {Vanzella}}]{Bakx_2023}
{Bakx}, T. J.~L.~C., {Zavala}, J.~A., {Mitsuhashi}, I., {et~al.} 2023, \mnras, 519, 5076, \dodoi{10.1093/mnras/stac3723}

\bibitem[{{Behroozi} {et~al.}(2019){Behroozi}, {Wechsler}, {Hearin}, \& {Conroy}}]{Behroozi_2019}
{Behroozi}, P., {Wechsler}, R.~H., {Hearin}, A.~P., \& {Conroy}, C. 2019, \mnras, 488, 3143, \dodoi{10.1093/mnras/stz1182}

\bibitem[{{Behroozi} {et~al.}(2013){Behroozi}, {Wechsler}, \& {Conroy}}]{Behroozi_2013}
{Behroozi}, P.~S., {Wechsler}, R.~H., \& {Conroy}, C. 2013, \apj, 770, 57, \dodoi{10.1088/0004-637X/770/1/57}

\bibitem[{{Biagetti} {et~al.}(2023){Biagetti}, {Franciolini}, \& {Riotto}}]{Biagetti_2023}
{Biagetti}, M., {Franciolini}, G., \& {Riotto}, A. 2023, \apj, 944, 113, \dodoi{10.3847/1538-4357/acb5ea}

\bibitem[{{Bird} {et~al.}(2023){Bird}, {Chang}, {Cui}, \& {Yang}}]{Chang_2023}
{Bird}, S., {Chang}, C.-F., {Cui}, Y., \& {Yang}, D. 2023, arXiv e-prints, arXiv:2307.10302.
\newblock \doarXiv{2307.10302}

\bibitem[{{Bouwens} {et~al.}(2021){Bouwens}, {Oesch}, {Stefanon}, {Illingworth}, {Labb{\'e}}, {Reddy}, {Atek}, {Montes}, {Naidu}, {Nanayakkara}, {Nelson}, \& {Wilkins}}]{Bouwens_2021}
{Bouwens}, R.~J., {Oesch}, P.~A., {Stefanon}, M., {et~al.} 2021, \aj, 162, 47, \dodoi{10.3847/1538-3881/abf83e}

\bibitem[{{Bouwens} {et~al.}(2023){Bouwens}, {Stefanon}, {Brammer}, {Oesch}, {Herard-Demanche}, {Illingworth}, {Matthee}, {Naidu}, {van Dokkum}, \& {van Leeuwen}}]{Bouwens_2023}
{Bouwens}, R.~J., {Stefanon}, M., {Brammer}, G., {et~al.} 2023, \mnras, 523, 1036, \dodoi{10.1093/mnras/stad1145}

\bibitem[{{Bowler} {et~al.}(2020){Bowler}, {Jarvis}, {Dunlop}, {McLure}, {McLeod}, {Adams}, {Milvang-Jensen}, \& {McCracken}}]{Bowler_2020}
{Bowler}, R.~A.~A., {Jarvis}, M.~J., {Dunlop}, J.~S., {et~al.} 2020, \mnras, 493, 2059, \dodoi{10.1093/mnras/staa313}

\bibitem[{{Boylan-Kolchin}(2023)}]{Boylan-Kolchin_2023}
{Boylan-Kolchin}, M. 2023, Nature Astronomy, \dodoi{10.1038/s41550-023-01937-7}

\bibitem[{{Byler} {et~al.}(2017){Byler}, {Dalcanton}, {Conroy}, \& {Johnson}}]{Byler_2017}
{Byler}, N., {Dalcanton}, J.~J., {Conroy}, C., \& {Johnson}, B.~D. 2017, \apj, 840, 44, \dodoi{10.3847/1538-4357/aa6c66}

\bibitem[{{Byrne} {et~al.}(2023){Byrne}, {Faucher-Gigu{\`e}re}, {Stern}, {Angl{\'e}s-Alc{\'a}zar}, {Wellons}, {Gurvich}, \& {Hopkins}}]{Byrne_2023}
{Byrne}, L., {Faucher-Gigu{\`e}re}, C.-A., {Stern}, J., {et~al.} 2023, \mnras, 520, 722, \dodoi{10.1093/mnras/stad171}

\bibitem[{{Castellano} {et~al.}(2023){Castellano}, {Fontana}, {Treu}, {Merlin}, {Santini}, {Bergamini}, {Grillo}, {Rosati}, {Acebron}, {Leethochawalit}, {Paris}, {Bonchi}, {Belfiori}, {Calabr{\`o}}, {Correnti}, {Nonino}, {Polenta}, {Trenti}, {Boyett}, {Brammer}, {Broadhurst}, {Caminha}, {Chen}, {Filippenko}, {Fortuni}, {Glazebrook}, {Mascia}, {Mason}, {Menci}, {Meneghetti}, {Mercurio}, {Metha}, {Morishita}, {Nanayakkara}, {Pentericci}, {Roberts-Borsani}, {Roy}, {Vanzella}, {Vulcani}, {Yang}, \& {Wang}}]{Castellano_2023}
{Castellano}, M., {Fontana}, A., {Treu}, T., {et~al.} 2023, \apjl, 948, L14, \dodoi{10.3847/2041-8213/accea5}

\bibitem[{{Cochrane} {et~al.}(2022){Cochrane}, {Hayward}, \& {Angl{\'e}s-Alc{\'a}zar}}]{Cochrane_2022}
{Cochrane}, R.~K., {Hayward}, C.~C., \& {Angl{\'e}s-Alc{\'a}zar}, D. 2022, \apjl, 939, L27, \dodoi{10.3847/2041-8213/ac951d}

\bibitem[{{Cochrane} {et~al.}(2019){Cochrane}, {Hayward}, {Angl{\'e}s-Alc{\'a}zar}, {Lotz}, {Parsotan}, {Ma}, {Kere{\v{s}}}, {Feldmann}, {Faucher-Gigu{\`e}re}, \& {Hopkins}}]{Cochrane_2019}
{Cochrane}, R.~K., {Hayward}, C.~C., {Angl{\'e}s-Alc{\'a}zar}, D., {et~al.} 2019, \mnras, 488, 1779, \dodoi{10.1093/mnras/stz1736}

\bibitem[{{Cullen} {et~al.}(2023){Cullen}, {McLure}, {McLeod}, {Dunlop}, {Donnan}, {Carnall}, {Bowler}, {Begley}, {Hamadouche}, \& {Stanton}}]{Cullen_2023}
{Cullen}, F., {McLure}, R.~J., {McLeod}, D.~J., {et~al.} 2023, \mnras, 520, 14, \dodoi{10.1093/mnras/stad073}

\bibitem[{{Curtis-Lake} {et~al.}(2023){Curtis-Lake}, {Carniani}, {Cameron}, {Charlot}, {Jakobsen}, {Maiolino}, {Bunker}, {Witstok}, {Smit}, {Chevallard}, {Willott}, {Ferruit}, {Arribas}, {Bonaventura}, {Curti}, {D'Eugenio}, {Franx}, {Giardino}, {Looser}, {L{\"u}tzgendorf}, {Maseda}, {Rawle}, {Rix}, {Rodr{\'\i}guez del Pino}, {{\"U}bler}, {Sirianni}, {Dressler}, {Egami}, {Eisenstein}, {Endsley}, {Hainline}, {Hausen}, {Johnson}, {Rieke}, {Robertson}, {Shivaei}, {Stark}, {Tacchella}, {Williams}, {Willmer}, {Bhatawdekar}, {Bowler}, {Boyett}, {Chen}, {de Graaff}, {Helton}, {Hviding}, {Jones}, {Kumari}, {Lyu}, {Nelson}, {Perna}, {Sandles}, {Saxena}, {Suess}, {Sun}, {Topping}, {Wallace}, \& {Whitler}}]{Curtis-Lake_2023NatAs}
{Curtis-Lake}, E., {Carniani}, S., {Cameron}, A., {et~al.} 2023, Nature Astronomy, 7, 622, \dodoi{10.1038/s41550-023-01918-w}

\bibitem[{{Dawoodbhoy} {et~al.}(2023){Dawoodbhoy}, {Shapiro}, {Ocvirk}, {Lewis}, {Aubert}, {Sorce}, {Ahn}, {Iliev}, {Park}, {Teyssier}, \& {Yepes}}]{Dawoodbhoy_2023}
{Dawoodbhoy}, T., {Shapiro}, P.~R., {Ocvirk}, P., {et~al.} 2023, arXiv e-prints, arXiv:2302.08523, \dodoi{10.48550/arXiv.2302.08523}

\bibitem[{{Dayal} \& {Giri}(2023)}]{DayalGiri_2023}
{Dayal}, P., \& {Giri}, S.~K. 2023, arXiv e-prints, arXiv:2303.14239, \dodoi{10.48550/arXiv.2303.14239}

\bibitem[{{Dekel} {et~al.}(2023){Dekel}, {Sarkar}, {Birnboim}, {Mandelker}, \& {Li}}]{Dekel_2023}
{Dekel}, A., {Sarkar}, K.~C., {Birnboim}, Y., {Mandelker}, N., \& {Li}, Z. 2023, \mnras, \dodoi{10.1093/mnras/stad1557}

\bibitem[{{Dom{\'\i}nguez} {et~al.}(2015){Dom{\'\i}nguez}, {Siana}, {Brooks}, {Christensen}, {Bruzual}, {Stark}, \& {Alavi}}]{Dominguez_2015}
{Dom{\'\i}nguez}, A., {Siana}, B., {Brooks}, A.~M., {et~al.} 2015, \mnras, 451, 839, \dodoi{10.1093/mnras/stv1001}

\bibitem[{{Donnan} {et~al.}(2023){Donnan}, {McLeod}, {Dunlop}, {McLure}, {Carnall}, {Begley}, {Cullen}, {Hamadouche}, {Bowler}, {Magee}, {McCracken}, {Milvang-Jensen}, {Moneti}, \& {Targett}}]{Donnan_2023}
{Donnan}, C.~T., {McLeod}, D.~J., {Dunlop}, J.~S., {et~al.} 2023, \mnras, 518, 6011, \dodoi{10.1093/mnras/stac3472}

\bibitem[{{Dressler} {et~al.}(2023){Dressler}, {Rieke}, {Eisenstein}, {Stark}, {Burns}, {Bhatawdekar}, {Bonaventura}, {Boyett}, {Bunker}, {Carniani}, {Charlot}, {Hausen}, {Misselt}, {Tacchella}, \& {Willmer}}]{Dressler_2023}
{Dressler}, A., {Rieke}, M., {Eisenstein}, D., {et~al.} 2023, arXiv e-prints, arXiv:2306.02469, \dodoi{10.48550/arXiv.2306.02469}

\bibitem[{{Eldridge} {et~al.}(2017){Eldridge}, {Stanway}, {Xiao}, {McClelland}, {Taylor}, {Ng}, {Greis}, \& {Bray}}]{Eldridge_2017}
{Eldridge}, J.~J., {Stanway}, E.~R., {Xiao}, L., {et~al.} 2017, \pasa, 34, e058, \dodoi{10.1017/pasa.2017.51}

\bibitem[{{Endsley} {et~al.}(2023){Endsley}, {Stark}, {Whitler}, {Topping}, {Johnson}, {Robertson}, {Tacchella}, {Alberts}, {Baker}, {Bhatawdekar}, {Boyett}, {Bunker}, {Cameron}, {Carniani}, {Charlot}, {Chen}, {Chevallard}, {Curtis-Lake}, {Danhaive}, {Egami}, {Eisenstein}, {Hainline}, {Helton}, {Ji}, {Looser}, {Maiolino}, {Nelson}, {Pusk{\'a}s}, {Rieke}, {Rieke}, {Rix}, {Sandles}, {Saxena}, {Simmonds}, {Smit}, {Sun}, {Williams}, {Willmer}, {Willott}, \& {Witstok}}]{Endsley_2023}
{Endsley}, R., {Stark}, D.~P., {Whitler}, L., {et~al.} 2023, arXiv e-prints, arXiv:2306.05295, \dodoi{10.48550/arXiv.2306.05295}

\bibitem[{{Faucher-Gigu{\`e}re}(2020)}]{FG20}
{Faucher-Gigu{\`e}re}, C.-A. 2020, \mnras, 493, 1614, \dodoi{10.1093/mnras/staa302}

\bibitem[{{Faucher-Gigu{\`e}re} {et~al.}(2009){Faucher-Gigu{\`e}re}, {Lidz}, {Zaldarriaga}, \& {Hernquist}}]{FG09}
{Faucher-Gigu{\`e}re}, C.-A., {Lidz}, A., {Zaldarriaga}, M., \& {Hernquist}, L. 2009, \apj, 703, 1416, \dodoi{10.1088/0004-637X/703/2/1416}

\bibitem[{{Ferrara} {et~al.}(2023){Ferrara}, {Pallottini}, \& {Dayal}}]{Ferrara_2023}
{Ferrara}, A., {Pallottini}, A., \& {Dayal}, P. 2023, \mnras, 522, 3986, \dodoi{10.1093/mnras/stad1095}

\bibitem[{{Finkelstein} {et~al.}(2022){Finkelstein}, {Bagley}, {Haro}, {Dickinson}, {Ferguson}, {Kartaltepe}, {Papovich}, {Burgarella}, {Kocevski}, {Huertas-Company}, {Iyer}, {Koekemoer}, {Larson}, {P{\'e}rez-Gonz{\'a}lez}, {Rose}, {Tacchella}, {Wilkins}, {Chworowsky}, {Medrano}, {Morales}, {Somerville}, {Yung}, {Fontana}, {Giavalisco}, {Grazian}, {Grogin}, {Kewley}, {Kirkpatrick}, {Kurczynski}, {Lotz}, {Pentericci}, {Pirzkal}, {Ravindranath}, {Ryan}, {Trump}, {Yang}, {Almaini}, {Amor{\'\i}n}, {Annunziatella}, {Backhaus}, {Barro}, {Behroozi}, {Bell}, {Bhatawdekar}, {Bisigello}, {Bromm}, {Buat}, {Buitrago}, {Calabr{\`o}}, {Casey}, {Castellano}, {Ch{\'a}vez Ortiz}, {Ciesla}, {Cleri}, {Cohen}, {Cole}, {Cooke}, {Cooper}, {Cooray}, {Costantin}, {Cox}, {Croton}, {Daddi}, {Dav{\'e}}, {de La Vega}, {Dekel}, {Elbaz}, {Estrada-Carpenter}, {Faber}, {Fern{\'a}ndez}, {Finkelstein}, {Freundlich}, {Fujimoto}, {Garc{\'\i}a-Argum{\'a}nez}, {Gardner}, {Gawiser}, {G{\'o}mez-Guijarro}, {Guo}, {Hamblin}, {Hamilton}, {Hathi},
  {Holwerda}, {Hirschmann}, {Hutchison}, {Jaskot}, {Jha}, {Jogee}, {Juneau}, {Jung}, {Kassin}, {Bail}, {Leung}, {Lucas}, {Magnelli}, {Mantha}, {Matharu}, {McGrath}, {McIntosh}, {Merlin}, {Mobasher}, {Newman}, {Nicholls}, {Pandya}, {Rafelski}, {Ronayne}, {Santini}, {Seill{\'e}}, {Shah}, {Shen}, {Simons}, {Snyder}, {Stanway}, {Straughn}, {Teplitz}, {Vanderhoof}, {Vega-Ferrero}, {Wang}, {Weiner}, {Willmer}, {Wuyts}, {Zavala}, \& {Ceers Team}}]{Finkelstein_2022}
{Finkelstein}, S.~L., {Bagley}, M.~B., {Haro}, P.~A., {et~al.} 2022, \apjl, 940, L55, \dodoi{10.3847/2041-8213/ac966e}

\bibitem[{{Flores Vel{\'a}zquez} {et~al.}(2021){Flores Vel{\'a}zquez}, {Gurvich}, {Faucher-Gigu{\`e}re}, {Bullock}, {Starkenburg}, {Moreno}, {Lazar}, {Mercado}, {Stern}, {Sparre}, {Hayward}, {Wetzel}, \& {El-Badry}}]{Flores_2021}
{Flores Vel{\'a}zquez}, J.~A., {Gurvich}, A.~B., {Faucher-Gigu{\`e}re}, C.-A., {et~al.} 2021, \mnras, 501, 4812, \dodoi{10.1093/mnras/staa3893}

\bibitem[{{Fudamoto} {et~al.}(2020){Fudamoto}, {Oesch}, {Faisst}, {B{\'e}thermin}, {Ginolfi}, {Khusanova}, {Loiacono}, {Le F{\`e}vre}, {Capak}, {Schaerer}, {Silverman}, {Cassata}, {Yan}, {Amorin}, {Bardelli}, {Boquien}, {Cimatti}, {Dessauges-Zavadsky}, {Fujimoto}, {Gruppioni}, {Hathi}, {Ibar}, {Jones}, {Koekemoer}, {Lagache}, {Lemaux}, {Maiolino}, {Narayanan}, {Pozzi}, {Riechers}, {Rodighiero}, {Talia}, {Toft}, {Vallini}, {Vergani}, {Zamorani}, \& {Zucca}}]{Fudamoto_2020}
{Fudamoto}, Y., {Oesch}, P.~A., {Faisst}, A., {et~al.} 2020, \aap, 643, A4, \dodoi{10.1051/0004-6361/202038163}

\bibitem[{{Fudamoto} {et~al.}(2021){Fudamoto}, {Oesch}, {Schouws}, {Stefanon}, {Smit}, {Bouwens}, {Bowler}, {Endsley}, {Gonzalez}, {Inami}, {Labbe}, {Stark}, {Aravena}, {Barrufet}, {da Cunha}, {Dayal}, {Ferrara}, {Graziani}, {Hodge}, {Hutter}, {Li}, {De Looze}, {Nanayakkara}, {Pallottini}, {Riechers}, {Schneider}, {Ucci}, {van der Werf}, \& {White}}]{Fudamoto_2021}
{Fudamoto}, Y., {Oesch}, P.~A., {Schouws}, S., {et~al.} 2021, \nat, 597, 489, \dodoi{10.1038/s41586-021-03846-z}

\bibitem[{{Furlanetto} \& {Mirocha}(2022)}]{FM_2022}
{Furlanetto}, S.~R., \& {Mirocha}, J. 2022, \mnras, 511, 3895, \dodoi{10.1093/mnras/stac310}

\bibitem[{{Furlanetto} \& {Mirocha}(2023)}]{FM_2023}
---. 2023, \mnras, 523, 5274, \dodoi{10.1093/mnras/stad1799}

\bibitem[{{Gnedin}(2000)}]{Gnedin_2000}
{Gnedin}, N.~Y. 2000, \apj, 542, 535, \dodoi{10.1086/317042}

\bibitem[{{Gong} {et~al.}(2023){Gong}, {Yue}, {Cao}, \& {Chen}}]{Gong_2023}
{Gong}, Y., {Yue}, B., {Cao}, Y., \& {Chen}, X. 2023, \apj, 947, 28, \dodoi{10.3847/1538-4357/acc109}

\bibitem[{{Grudi{\'c}} {et~al.}(2018){Grudi{\'c}}, {Hopkins}, {Faucher-Gigu{\`e}re}, {Quataert}, {Murray}, \& {Kere{\v{s}}}}]{Grudic_2018}
{Grudi{\'c}}, M.~Y., {Hopkins}, P.~F., {Faucher-Gigu{\`e}re}, C.-A., {et~al.} 2018, \mnras, 475, 3511, \dodoi{10.1093/mnras/sty035}

\bibitem[{{Gurvich} {et~al.}(2020){Gurvich}, {Faucher-Gigu{\`e}re}, {Richings}, {Hopkins}, {Grudi{\'c}}, {Hafen}, {Wellons}, {Stern}, {Quataert}, {Chan}, {Orr}, {Kere{\v{s}}}, {Wetzel}, {Hayward}, {Loebman}, \& {Murray}}]{Gurvich_2020}
{Gurvich}, A.~B., {Faucher-Gigu{\`e}re}, C.-A., {Richings}, A.~J., {et~al.} 2020, \mnras, 498, 3664, \dodoi{10.1093/mnras/staa2578}

\bibitem[{{Gurvich} {et~al.}(2023){Gurvich}, {Stern}, {Faucher-Gigu{\`e}re}, {Hopkins}, {Wetzel}, {Moreno}, {Hayward}, {Richings}, \& {Hafen}}]{Gurvich_2023}
{Gurvich}, A.~B., {Stern}, J., {Faucher-Gigu{\`e}re}, C.-A., {et~al.} 2023, \mnras, 519, 2598, \dodoi{10.1093/mnras/stac3712}

\bibitem[{{Harikane} {et~al.}(2023{\natexlab{a}}){Harikane}, {Nakajima}, {Ouchi}, {Umeda}, {Isobe}, {Ono}, {Xu}, \& {Zhang}}]{Harikane_2023arXiv}
{Harikane}, Y., {Nakajima}, K., {Ouchi}, M., {et~al.} 2023{\natexlab{a}}, arXiv e-prints, arXiv:2304.06658, \dodoi{10.48550/arXiv.2304.06658}

\bibitem[{{Harikane} {et~al.}(2018){Harikane}, {Ouchi}, {Ono}, {Saito}, {Behroozi}, {More}, {Shimasaku}, {Toshikawa}, {Lin}, {Akiyama}, {Coupon}, {Komiyama}, {Konno}, {Lin}, {Miyazaki}, {Nishizawa}, {Shibuya}, \& {Silverman}}]{Harikane_2018}
{Harikane}, Y., {Ouchi}, M., {Ono}, Y., {et~al.} 2018, \pasj, 70, S11, \dodoi{10.1093/pasj/psx097}

\bibitem[{{Harikane} {et~al.}(2023{\natexlab{b}}){Harikane}, {Ouchi}, {Oguri}, {Ono}, {Nakajima}, {Isobe}, {Umeda}, {Mawatari}, \& {Zhang}}]{Harikane_2023ApJS}
{Harikane}, Y., {Ouchi}, M., {Oguri}, M., {et~al.} 2023{\natexlab{b}}, \apjs, 265, 5, \dodoi{10.3847/1538-4365/acaaa9}

\bibitem[{{Harikane} {et~al.}(2023{\natexlab{c}}){Harikane}, {Zhang}, {Nakajima}, {Ouchi}, {Isobe}, {Ono}, {Hatano}, {Xu}, \& {Umeda}}]{Harikane_2023AGN}
{Harikane}, Y., {Zhang}, Y., {Nakajima}, K., {et~al.} 2023{\natexlab{c}}, arXiv e-prints, arXiv:2303.11946, \dodoi{10.48550/arXiv.2303.11946}

\bibitem[{{Hassan} {et~al.}(2023){Hassan}, {Lovell}, {Madau}, {Huertas-Company}, {Somerville}, {Burkhart}, {Dixon}, {Feldmann}, {Starkenburg}, {Wu}, {Kragh Jespersen}, {Gelfand}, \& {Bera}}]{Hassan_2023}
{Hassan}, S., {Lovell}, C.~C., {Madau}, P., {et~al.} 2023, arXiv e-prints, arXiv:2305.02703, \dodoi{10.48550/arXiv.2305.02703}

\bibitem[{{Hirano} \& {Yoshida}(2023)}]{HiranoYoshida_2023}
{Hirano}, S., \& {Yoshida}, N. 2023, arXiv e-prints, arXiv:2306.11993, \dodoi{10.48550/arXiv.2306.11993}

\bibitem[{{Hopkins}(2015)}]{Hopkins_2015_GIZMO}
{Hopkins}, P.~F. 2015, \mnras, 450, 53, \dodoi{10.1093/mnras/stv195}

\bibitem[{{Hopkins} {et~al.}(2014){Hopkins}, {Kere{\v{s}}}, {O{\~n}orbe}, {Faucher-Gigu{\`e}re}, {Quataert}, {Murray}, \& {Bullock}}]{Hopkins_2014}
{Hopkins}, P.~F., {Kere{\v{s}}}, D., {O{\~n}orbe}, J., {et~al.} 2014, \mnras, 445, 581, \dodoi{10.1093/mnras/stu1738}

\bibitem[{{Hopkins} {et~al.}(2018){Hopkins}, {Wetzel}, {Kere{\v{s}}}, {Faucher-Gigu{\`e}re}, {Quataert}, {Boylan-Kolchin}, {Murray}, {Hayward}, {Garrison-Kimmel}, {Hummels}, {Feldmann}, {Torrey}, {Ma}, {Angl{\'e}s-Alc{\'a}zar}, {Su}, {Orr}, {Schmitz}, {Escala}, {Sanderson}, {Grudi{\'c}}, {Hafen}, {Kim}, {Fitts}, {Bullock}, {Wheeler}, {Chan}, {Elbert}, \& {Narayanan}}]{Hopkins_2018}
{Hopkins}, P.~F., {Wetzel}, A., {Kere{\v{s}}}, D., {et~al.} 2018, \mnras, 480, 800, \dodoi{10.1093/mnras/sty1690}

\bibitem[{{Hopkins} {et~al.}(2023){Hopkins}, {Gurvich}, {Shen}, {Hafen}, {Grudi{\'c}}, {Kurinchi-Vendhan}, {Hayward}, {Jiang}, {Orr}, {Wetzel}, {Kere{\v{s}}}, {Stern}, {Faucher-Gigu{\'e}re}, {Bullock}, {Wheeler}, {El-Badry}, {Loebman}, {Moreno}, {Boylan-Kolchin}, \& {Quataert}}]{Hopkins_2023}
{Hopkins}, P.~F., {Gurvich}, A.~B., {Shen}, X., {et~al.} 2023, \mnras, \dodoi{10.1093/mnras/stad1902}

\bibitem[{{Hu} {et~al.}(2023){Hu}, {Smith}, {Teyssier}, {Bryan}, {Verbeke}, {Emerick}, {Somerville}, {Burkhart}, {Li}, {Forbes}, \& {Starkenburg}}]{Hu_2023}
{Hu}, C.-Y., {Smith}, M.~C., {Teyssier}, R., {et~al.} 2023, \apj, 950, 132, \dodoi{10.3847/1538-4357/accf9e}

\bibitem[{{Inayoshi} {et~al.}(2022){Inayoshi}, {Harikane}, {Inoue}, {Li}, \& {Ho}}]{Inayoshi_2022}
{Inayoshi}, K., {Harikane}, Y., {Inoue}, A.~K., {Li}, W., \& {Ho}, L.~C. 2022, \apjl, 938, L10, \dodoi{10.3847/2041-8213/ac9310}

\bibitem[{{Iyer} {et~al.}(2020){Iyer}, {Tacchella}, {Genel}, {Hayward}, {Hernquist}, {Brooks}, {Caplar}, {Dav{\'e}}, {Diemer}, {Forbes}, {Gawiser}, {Somerville}, \& {Starkenburg}}]{Iyer_2020}
{Iyer}, K.~G., {Tacchella}, S., {Genel}, S., {et~al.} 2020, \mnras, 498, 430, \dodoi{10.1093/mnras/staa2150}

\bibitem[{{Kannan} {et~al.}(2023){Kannan}, {Springel}, {Hernquist}, {Pakmor}, {Delgado}, {Hadzhiyska}, {Hern{\'a}ndez-Aguayo}, {Barrera}, {Ferlito}, {Bose}, {White}, {Frenk}, {Smith}, \& {Garaldi}}]{Kannan_2023}
{Kannan}, R., {Springel}, V., {Hernquist}, L., {et~al.} 2023, \mnras, 524, 2594, \dodoi{10.1093/mnras/stac3743}

\bibitem[{{Keller} {et~al.}(2023){Keller}, {Munshi}, {Trebitsch}, \& {Tremmel}}]{Keller_2023}
{Keller}, B.~W., {Munshi}, F., {Trebitsch}, M., \& {Tremmel}, M. 2023, \apjl, 943, L28, \dodoi{10.3847/2041-8213/acb148}

\bibitem[{{Knollmann} \& {Knebe}(2009)}]{KnollmannKnebe_2009}
{Knollmann}, S.~R., \& {Knebe}, A. 2009, \apjs, 182, 608, \dodoi{10.1088/0067-0049/182/2/608}

\bibitem[{{Kroupa}(2001)}]{Kroupa_2001}
{Kroupa}, P. 2001, \mnras, 322, 231, \dodoi{10.1046/j.1365-8711.2001.04022.x}

\bibitem[{{Leethochawalit} {et~al.}(2022){Leethochawalit}, {Roberts-Borsani}, {Morishita}, {Trenti}, \& {Treu}}]{Leethochawalit_2022}
{Leethochawalit}, N., {Roberts-Borsani}, G., {Morishita}, T., {Trenti}, M., \& {Treu}, T. 2022, arXiv e-prints, arXiv:2205.15388, \dodoi{10.48550/arXiv.2205.15388}

\bibitem[{{Looser} {et~al.}(2023{\natexlab{a}}){Looser}, {D'Eugenio}, {Maiolino}, {Witstok}, {Sandles}, {Curtis-Lake}, {Chevallard}, {Tacchella}, {Johnson}, {Baker}, {Suess}, {Carniani}, {Ferruit}, {Arribas}, {Bonaventura}, {Bunker}, {Cameron}, {Charlot}, {Curti}, {de Graaff}, {Maseda}, {Rawle}, {Rix}, {Rodriguez Del Pino}, {Smit}, {{\"U}bler}, {Willott}, {Alberts}, {Egami}, {Eisenstein}, {Endsley}, {Hausen}, {Rieke}, {Robertson}, {Shivaei}, {Williams}, {Boyett}, {Chen}, {Ji}, {Jones}, {Kumari}, {Nelson}, {Perna}, {Saxena}, \& {Scholtz}}]{Looser_2023QG}
{Looser}, T.~J., {D'Eugenio}, F., {Maiolino}, R., {et~al.} 2023{\natexlab{a}}, arXiv e-prints, arXiv:2302.14155, \dodoi{10.48550/arXiv.2302.14155}

\bibitem[{{Looser} {et~al.}(2023{\natexlab{b}}){Looser}, {D'Eugenio}, {Maiolino}, {Tacchella}, {Curti}, {Arribas}, {Baker}, {Baum}, {Bonaventura}, {Boyett}, {Bunker}, {Carniani}, {Charlot}, {Chevallard}, {Curtis-Lake}, {Danhaive}, {Eisenstein}, {de Graaff}, {Hainline}, {Ji}, {Johnson}, {Kumari}, {Nelson}, {Parlanti}, {Rix}, {Robertson}, {Rodr{\'\i}guez Del Pino}, {Sandles}, {Scholtz}, {Smit}, {Stark}, {{\"U}bler}, {Williams}, {Willott}, \& {Witstok}}]{Looser_2023}
---. 2023{\natexlab{b}}, arXiv e-prints, arXiv:2306.02470, \dodoi{10.48550/arXiv.2306.02470}

\bibitem[{{Ma} {et~al.}(2018{\natexlab{a}}){Ma}, {Hopkins}, {Boylan-Kolchin}, {Faucher-Gigu{\`e}re}, {Quataert}, {Feldmann}, {Garrison-Kimmel}, {Hayward}, {Kere{\v{s}}}, \& {Wetzel}}]{Ma_2018_size}
{Ma}, X., {Hopkins}, P.~F., {Boylan-Kolchin}, M., {et~al.} 2018{\natexlab{a}}, \mnras, 477, 219, \dodoi{10.1093/mnras/sty684}

\bibitem[{{Ma} {et~al.}(2018{\natexlab{b}}){Ma}, {Hopkins}, {Garrison-Kimmel}, {Faucher-Gigu{\`e}re}, {Quataert}, {Boylan-Kolchin}, {Hayward}, {Feldmann}, \& {Kere{\v{s}}}}]{Ma_2018_lf}
{Ma}, X., {Hopkins}, P.~F., {Garrison-Kimmel}, S., {et~al.} 2018{\natexlab{b}}, \mnras, 478, 1694, \dodoi{10.1093/mnras/sty1024}

\bibitem[{{Ma} {et~al.}(2019){Ma}, {Hayward}, {Casey}, {Hopkins}, {Quataert}, {Liang}, {Faucher-Gigu{\`e}re}, {Feldmann}, \& {Kere{\v{s}}}}]{Ma_2019}
{Ma}, X., {Hayward}, C.~C., {Casey}, C.~M., {et~al.} 2019, \mnras, 487, 1844, \dodoi{10.1093/mnras/stz1324}

\bibitem[{{Mason} {et~al.}(2015){Mason}, {Trenti}, \& {Treu}}]{Mason_2015}
{Mason}, C.~A., {Trenti}, M., \& {Treu}, T. 2015, \apj, 813, 21, \dodoi{10.1088/0004-637X/813/1/21}

\bibitem[{{Mason} {et~al.}(2023){Mason}, {Trenti}, \& {Treu}}]{Mason_2023}
---. 2023, \mnras, 521, 497, \dodoi{10.1093/mnras/stad035}

\bibitem[{{McCaffrey} {et~al.}(2023){McCaffrey}, {Hardin}, {Wise}, \& {Regan}}]{McCaffrey_2023}
{McCaffrey}, J., {Hardin}, S., {Wise}, J., \& {Regan}, J. 2023, arXiv e-prints, arXiv:2304.13755, \dodoi{10.48550/arXiv.2304.13755}

\bibitem[{{McLeod} {et~al.}(2023){McLeod}, {Donnan}, {McLure}, {Dunlop}, {Magee}, {Begley}, {Carnall}, {Cullen}, {Ellis}, {Hamadouche}, \& {Stanton}}]{McLeod_2023}
{McLeod}, D.~J., {Donnan}, C.~T., {McLure}, R.~J., {et~al.} 2023, arXiv e-prints, arXiv:2304.14469, \dodoi{10.48550/arXiv.2304.14469}

\bibitem[{{Melia}(2023)}]{Melia_2023}
{Melia}, F. 2023, \mnras, 521, L85, \dodoi{10.1093/mnrasl/slad025}

\bibitem[{{Mirocha} \& {Furlanetto}(2023)}]{MF_2023}
{Mirocha}, J., \& {Furlanetto}, S.~R. 2023, \mnras, 519, 843, \dodoi{10.1093/mnras/stac3578}

\bibitem[{{Mu{\~n}oz} {et~al.}(2023){Mu{\~n}oz}, {Mirocha}, {Furlanetto}, \& {Sabti}}]{Munoz_2023}
{Mu{\~n}oz}, J.~B., {Mirocha}, J., {Furlanetto}, S., \& {Sabti}, N. 2023, arXiv e-prints, arXiv:2306.09403, \dodoi{10.48550/arXiv.2306.09403}

\bibitem[{{Muratov} {et~al.}(2015){Muratov}, {Kere{\v{s}}}, {Faucher-Gigu{\`e}re}, {Hopkins}, {Quataert}, \& {Murray}}]{Muratov_2015}
{Muratov}, A.~L., {Kere{\v{s}}}, D., {Faucher-Gigu{\`e}re}, C.-A., {et~al.} 2015, \mnras, 454, 2691, \dodoi{10.1093/mnras/stv2126}

\bibitem[{{Murray} {et~al.}(2013){Murray}, {Power}, \& {Robotham}}]{Murray_2013}
{Murray}, S.~G., {Power}, C., \& {Robotham}, A.~S.~G. 2013, Astronomy and Computing, 3, 23, \dodoi{10.1016/j.ascom.2013.11.001}

\bibitem[{{Naidu} {et~al.}(2022{\natexlab{a}}){Naidu}, {Oesch}, {van Dokkum}, {Nelson}, {Suess}, {Brammer}, {Whitaker}, {Illingworth}, {Bouwens}, {Tacchella}, {Matthee}, {Allen}, {Bezanson}, {Conroy}, {Labbe}, {Leja}, {Leonova}, {Magee}, {Price}, {Setton}, {Strait}, {Stefanon}, {Toft}, {Weaver}, \& {Weibel}}]{Naidu_2022}
{Naidu}, R.~P., {Oesch}, P.~A., {van Dokkum}, P., {et~al.} 2022{\natexlab{a}}, \apjl, 940, L14, \dodoi{10.3847/2041-8213/ac9b22}

\bibitem[{{Naidu} {et~al.}(2022{\natexlab{b}}){Naidu}, {Oesch}, {Setton}, {Matthee}, {Conroy}, {Johnson}, {Weaver}, {Bouwens}, {Brammer}, {Dayal}, {Illingworth}, {Barrufet}, {Belli}, {Bezanson}, {Bose}, {Heintz}, {Leja}, {Leonova}, {Marques-Chaves}, {Stefanon}, {Toft}, {van der Wel}, {van Dokkum}, {Weibel}, \& {Whitaker}}]{Naidu_2022arXiv}
{Naidu}, R.~P., {Oesch}, P.~A., {Setton}, D.~J., {et~al.} 2022{\natexlab{b}}, arXiv e-prints, arXiv:2208.02794, \dodoi{10.48550/arXiv.2208.02794}

\bibitem[{{Noh} \& {McQuinn}(2014)}]{Noh_McQuinn_2014}
{Noh}, Y., \& {McQuinn}, M. 2014, \mnras, 444, 503, \dodoi{10.1093/mnras/stu1412}

\bibitem[{{Ocvirk} {et~al.}(2020){Ocvirk}, {Aubert}, {Sorce}, {Shapiro}, {Deparis}, {Dawoodbhoy}, {Lewis}, {Teyssier}, {Yepes}, {Gottl{\"o}ber}, {Ahn}, {Iliev}, \& {Hoffman}}]{Ocvirk_2020}
{Ocvirk}, P., {Aubert}, D., {Sorce}, J.~G., {et~al.} 2020, \mnras, 496, 4087, \dodoi{10.1093/mnras/staa1266}

\bibitem[{{Oesch} {et~al.}(2018){Oesch}, {Bouwens}, {Illingworth}, {Labb{\'e}}, \& {Stefanon}}]{Oesch_2018}
{Oesch}, P.~A., {Bouwens}, R.~J., {Illingworth}, G.~D., {Labb{\'e}}, I., \& {Stefanon}, M. 2018, \apj, 855, 105, \dodoi{10.3847/1538-4357/aab03f}

\bibitem[{{Oke} \& {Gunn}(1983)}]{OG_1983}
{Oke}, J.~B., \& {Gunn}, J.~E. 1983, \apj, 266, 713, \dodoi{10.1086/160817}

\bibitem[{{Orr} {et~al.}(2018){Orr}, {Hayward}, {Hopkins}, {Chan}, {Faucher-Gigu{\`e}re}, {Feldmann}, {Kere{\v{s}}}, {Murray}, \& {Quataert}}]{Orr_2018}
{Orr}, M.~E., {Hayward}, C.~C., {Hopkins}, P.~F., {et~al.} 2018, \mnras, 478, 3653, \dodoi{10.1093/mnras/sty1241}

\bibitem[{{Padmanabhan} \& {Loeb}(2023)}]{Padmanabhan_2023}
{Padmanabhan}, H., \& {Loeb}, A. 2023, arXiv e-prints, arXiv:2306.04684, \dodoi{10.48550/arXiv.2306.04684}

\bibitem[{{Pallottini} \& {Ferrara}(2023)}]{PallottiniFerrara_2023}
{Pallottini}, A., \& {Ferrara}, A. 2023, arXiv e-prints, arXiv:2307.03219, \dodoi{10.48550/arXiv.2307.03219}

\bibitem[{{Pallottini} {et~al.}(2022){Pallottini}, {Ferrara}, {Gallerani}, {Behrens}, {Kohandel}, {Carniani}, {Vallini}, {Salvadori}, {Gelli}, {Sommovigo}, {D'Odorico}, {Di Mascia}, \& {Pizzati}}]{Pallottini_2022}
{Pallottini}, A., {Ferrara}, A., {Gallerani}, S., {et~al.} 2022, \mnras, 513, 5621, \dodoi{10.1093/mnras/stac1281}

\bibitem[{{Parashari} \& {Laha}(2023)}]{ParashariLaha_2023}
{Parashari}, P., \& {Laha}, R. 2023, arXiv e-prints, arXiv:2305.00999, \dodoi{10.48550/arXiv.2305.00999}

\bibitem[{{P{\'e}rez-Gonz{\'a}lez} {et~al.}(2023){P{\'e}rez-Gonz{\'a}lez}, {Costantin}, {Langeroodi}, {Rinaldi}, {Annunziatella}, {Ilbert}, {Colina}, {Noorgaard-Nielsen}, {Greve}, {Ostlin}, {Wright}, {Alonso-Herrero}, {{\'A}lvarez-M{\'a}rquez}, {Caputi}, {Eckart}, {Le F{\`e}vre}, {Labiano}, {Garc{\'\i}a-Mar{\'\i}n}, {Hjorth}, {Kendrew}, {Pye}, {Tikkanen}, {van der Werf}, {Walter}, {Ward}, {Bik}, {Boogaard}, {Bosman}, {Crespo G{\'o}mez}, {Gillman}, {Iani}, {Jermann}, {Melinder}, {Meyer}, {Moutard}, {van Dishoek}, {Henning}, {Lagage}, {Guedel}, {Peissker}, {Ray}, {Vandenbussche}, {Garc{\'\i}a-Argum{\'a}nez}, \& {Mar{\'\i}a M{\'e}rida}}]{Perez-Gonzalez_2023}
{P{\'e}rez-Gonz{\'a}lez}, P.~G., {Costantin}, L., {Langeroodi}, D., {et~al.} 2023, arXiv e-prints, arXiv:2302.02429, \dodoi{10.48550/arXiv.2302.02429}

\bibitem[{{Planck Collaboration} {et~al.}(2020){Planck Collaboration}, {Aghanim}, {Akrami}, {Ashdown}, {Aumont}, {Baccigalupi}, {Ballardini}, {Banday}, {Barreiro}, {Bartolo}, {Basak}, {Battye}, {Benabed}, {Bernard}, {Bersanelli}, {Bielewicz}, {Bock}, {Bond}, {Borrill}, {Bouchet}, {Boulanger}, {Bucher}, {Burigana}, {Butler}, {Calabrese}, {Cardoso}, {Carron}, {Challinor}, {Chiang}, {Chluba}, {Colombo}, {Combet}, {Contreras}, {Crill}, {Cuttaia}, {de Bernardis}, {de Zotti}, {Delabrouille}, {Delouis}, {Di Valentino}, {Diego}, {Dor{\'e}}, {Douspis}, {Ducout}, {Dupac}, {Dusini}, {Efstathiou}, {Elsner}, {En{\ss}lin}, {Eriksen}, {Fantaye}, {Farhang}, {Fergusson}, {Fernandez-Cobos}, {Finelli}, {Forastieri}, {Frailis}, {Fraisse}, {Franceschi}, {Frolov}, {Galeotta}, {Galli}, {Ganga}, {G{\'e}nova-Santos}, {Gerbino}, {Ghosh}, {Gonz{\'a}lez-Nuevo}, {G{\'o}rski}, {Gratton}, {Gruppuso}, {Gudmundsson}, {Hamann}, {Handley}, {Hansen}, {Herranz}, {Hildebrandt}, {Hivon}, {Huang}, {Jaffe}, {Jones}, {Karakci}, {Keih{\"a}nen},
  {Keskitalo}, {Kiiveri}, {Kim}, {Kisner}, {Knox}, {Krachmalnicoff}, {Kunz}, {Kurki-Suonio}, {Lagache}, {Lamarre}, {Lasenby}, {Lattanzi}, {Lawrence}, {Le Jeune}, {Lemos}, {Lesgourgues}, {Levrier}, {Lewis}, {Liguori}, {Lilje}, {Lilley}, {Lindholm}, {L{\'o}pez-Caniego}, {Lubin}, {Ma}, {Mac{\'\i}as-P{\'e}rez}, {Maggio}, {Maino}, {Mandolesi}, {Mangilli}, {Marcos-Caballero}, {Maris}, {Martin}, {Martinelli}, {Mart{\'\i}nez-Gonz{\'a}lez}, {Matarrese}, {Mauri}, {McEwen}, {Meinhold}, {Melchiorri}, {Mennella}, {Migliaccio}, {Millea}, {Mitra}, {Miville-Desch{\^e}nes}, {Molinari}, {Montier}, {Morgante}, {Moss}, {Natoli}, {N{\o}rgaard-Nielsen}, {Pagano}, {Paoletti}, {Partridge}, {Patanchon}, {Peiris}, {Perrotta}, {Pettorino}, {Piacentini}, {Polastri}, {Polenta}, {Puget}, {Rachen}, {Reinecke}, {Remazeilles}, {Renzi}, {Rocha}, {Rosset}, {Roudier}, {Rubi{\~n}o-Mart{\'\i}n}, {Ruiz-Granados}, {Salvati}, {Sandri}, {Savelainen}, {Scott}, {Shellard}, {Sirignano}, {Sirri}, {Spencer}, {Sunyaev}, {Suur-Uski}, {Tauber}, {Tavagnacco},
  {Tenti}, {Toffolatti}, {Tomasi}, {Trombetti}, {Valenziano}, {Valiviita}, {Van Tent}, {Vibert}, {Vielva}, {Villa}, {Vittorio}, {Wandelt}, {Wehus}, {White}, {White}, {Zacchei}, \& {Zonca}}]{Planck_2018}
{Planck Collaboration}, {Aghanim}, N., {Akrami}, Y., {et~al.} 2020, \aap, 641, A6, \dodoi{10.1051/0004-6361/201833910}

\bibitem[{{Qin} {et~al.}(2023){Qin}, {Balu}, \& {Wyithe}}]{Qin_2023}
{Qin}, Y., {Balu}, S., \& {Wyithe}, J. S.~B. 2023, arXiv e-prints, arXiv:2305.17959, \dodoi{10.48550/arXiv.2305.17959}

\bibitem[{{Reddy} {et~al.}(2018){Reddy}, {Oesch}, {Bouwens}, {Montes}, {Illingworth}, {Steidel}, {van Dokkum}, {Atek}, {Carollo}, {Cibinel}, {Holden}, {Labb{\'e}}, {Magee}, {Morselli}, {Nelson}, \& {Wilkins}}]{Reddy_2018}
{Reddy}, N.~A., {Oesch}, P.~A., {Bouwens}, R.~J., {et~al.} 2018, \apj, 853, 56, \dodoi{10.3847/1538-4357/aaa3e7}

\bibitem[{{Robertson}(2022)}]{Robertson_2022ARA&A}
{Robertson}, B.~E. 2022, \araa, 60, 121, \dodoi{10.1146/annurev-astro-120221-044656}

\bibitem[{{Robertson} {et~al.}(2023){Robertson}, {Tacchella}, {Johnson}, {Hainline}, {Whitler}, {Eisenstein}, {Endsley}, {Rieke}, {Stark}, {Alberts}, {Dressler}, {Egami}, {Hausen}, {Rieke}, {Shivaei}, {Williams}, {Willmer}, {Arribas}, {Bonaventura}, {Bunker}, {Cameron}, {Carniani}, {Charlot}, {Chevallard}, {Curti}, {Curtis-Lake}, {D'Eugenio}, {Jakobsen}, {Looser}, {L{\"u}tzgendorf}, {Maiolino}, {Maseda}, {Rawle}, {Rix}, {Smit}, {{\"U}bler}, {Willott}, {Witstok}, {Baum}, {Bhatawdekar}, {Boyett}, {Chen}, {de Graaff}, {Florian}, {Helton}, {Hviding}, {Ji}, {Kumari}, {Lyu}, {Nelson}, {Sandles}, {Saxena}, {Suess}, {Sun}, {Topping}, \& {Wallace}}]{Robertson_2023NatAs}
{Robertson}, B.~E., {Tacchella}, S., {Johnson}, B.~D., {et~al.} 2023, Nature Astronomy, 7, 611, \dodoi{10.1038/s41550-023-01921-1}

\bibitem[{{Rojas-Ruiz} {et~al.}(2020){Rojas-Ruiz}, {Finkelstein}, {Bagley}, {Stevans}, {Finkelstein}, {Larson}, {Mechtley}, \& {Diekmann}}]{Rojas-Ruiz_2020}
{Rojas-Ruiz}, S., {Finkelstein}, S.~L., {Bagley}, M.~B., {et~al.} 2020, \apj, 891, 146, \dodoi{10.3847/1538-4357/ab7659}

\bibitem[{{Sabti} {et~al.}(2023){Sabti}, {Mu{\~n}oz}, \& {Kamionkowski}}]{Sabti_2023}
{Sabti}, N., {Mu{\~n}oz}, J.~B., \& {Kamionkowski}, M. 2023, arXiv e-prints, arXiv:2305.07049, \dodoi{10.48550/arXiv.2305.07049}

\bibitem[{{Shen} {et~al.}(2023){Shen}, {Vogelsberger}, {Boylan-Kolchin}, {Tacchella}, \& {Kannan}}]{Shen_2023}
{Shen}, X., {Vogelsberger}, M., {Boylan-Kolchin}, M., {Tacchella}, S., \& {Kannan}, R. 2023, arXiv e-prints, arXiv:2305.05679, \dodoi{10.48550/arXiv.2305.05679}

\bibitem[{{Shen} {et~al.}(2022){Shen}, {Vogelsberger}, {Nelson}, {Tacchella}, {Hernquist}, {Springel}, {Marinacci}, \& {Torrey}}]{Shen_2022}
{Shen}, X., {Vogelsberger}, M., {Nelson}, D., {et~al.} 2022, \mnras, 510, 5560, \dodoi{10.1093/mnras/stab3794}

\bibitem[{{Sipple} \& {Lidz}(2023)}]{SippleLidz_2023}
{Sipple}, J., \& {Lidz}, A. 2023, arXiv e-prints, arXiv:2306.12087, \dodoi{10.48550/arXiv.2306.12087}

\bibitem[{{Sparre} {et~al.}(2017){Sparre}, {Hayward}, {Feldmann}, {Faucher-Gigu{\`e}re}, {Muratov}, {Kere{\v{s}}}, \& {Hopkins}}]{Sparre_2017}
{Sparre}, M., {Hayward}, C.~C., {Feldmann}, R., {et~al.} 2017, \mnras, 466, 88, \dodoi{10.1093/mnras/stw3011}

\bibitem[{{Stern} {et~al.}(2021){Stern}, {Faucher-Gigu{\`e}re}, {Fielding}, {Quataert}, {Hafen}, {Gurvich}, {Ma}, {Byrne}, {El-Badry}, {Angl{\'e}s-Alc{\'a}zar}, {Chan}, {Feldmann}, {Kere{\v{s}}}, {Wetzel}, {Murray}, \& {Hopkins}}]{Stern_2021_ICV}
{Stern}, J., {Faucher-Gigu{\`e}re}, C.-A., {Fielding}, D., {et~al.} 2021, \apj, 911, 88, \dodoi{10.3847/1538-4357/abd776}

\bibitem[{{Sun} {et~al.}(2023{\natexlab{a}}){Sun}, {Faucher-Gigu{\`e}re}, {Hayward}, \& {Shen}}]{Sun_2023}
{Sun}, G., {Faucher-Gigu{\`e}re}, C.-A., {Hayward}, C.~C., \& {Shen}, X. 2023{\natexlab{a}}, arXiv e-prints, arXiv:2305.02713, \dodoi{10.48550/arXiv.2305.02713}

\bibitem[{{Sun} \& {Furlanetto}(2016)}]{SF_2016}
{Sun}, G., \& {Furlanetto}, S.~R. 2016, \mnras, 460, 417, \dodoi{10.1093/mnras/stw980}

\bibitem[{{Sun} {et~al.}(2023{\natexlab{b}}){Sun}, {Lidz}, {Faisst}, \& {Faucher-Gigu{\`e}re}}]{Sun_2023_EBL}
{Sun}, G., {Lidz}, A., {Faisst}, A.~L., \& {Faucher-Gigu{\`e}re}, C.-A. 2023{\natexlab{b}}, \mnras, \dodoi{10.1093/mnras/stad2000}

\bibitem[{{Tacchella} {et~al.}(2018){Tacchella}, {Bose}, {Conroy}, {Eisenstein}, \& {Johnson}}]{Tacchella_2018}
{Tacchella}, S., {Bose}, S., {Conroy}, C., {Eisenstein}, D.~J., \& {Johnson}, B.~D. 2018, \apj, 868, 92, \dodoi{10.3847/1538-4357/aae8e0}

\bibitem[{{Tacchella} {et~al.}(2020){Tacchella}, {Forbes}, \& {Caplar}}]{Tacchella_2020}
{Tacchella}, S., {Forbes}, J.~C., \& {Caplar}, N. 2020, \mnras, 497, 698, \dodoi{10.1093/mnras/staa1838}

\bibitem[{{Tamura} {et~al.}(2019){Tamura}, {Mawatari}, {Hashimoto}, {Inoue}, {Zackrisson}, {Christensen}, {Binggeli}, {Matsuda}, {Matsuo}, {Takeuchi}, {Asano}, {Sunaga}, {Shimizu}, {Okamoto}, {Yoshida}, {Lee}, {Shibuya}, {Taniguchi}, {Umehata}, {Hatsukade}, {Kohno}, \& {Ota}}]{Tamura_2019}
{Tamura}, Y., {Mawatari}, K., {Hashimoto}, T., {et~al.} 2019, \apj, 874, 27, \dodoi{10.3847/1538-4357/ab0374}

\bibitem[{{Topping} {et~al.}(2023){Topping}, {Stark}, {Endsley}, {Whitler}, {Hainline}, {Johnson}, {Robertson}, {Tacchella}, {Chen}, {Alberts}, {Baker}, {Bunker}, {Carniani}, {Charlot}, {Chevallard}, {Curtis-Lake}, {DeCoursey}, {Egami}, {Eisenstein}, {Ji}, {Maiolino}, {Williams}, {Willmer}, {Willott}, \& {Witstok}}]{Topping_2023}
{Topping}, M.~W., {Stark}, D.~P., {Endsley}, R., {et~al.} 2023, arXiv e-prints, arXiv:2307.08835, \dodoi{10.48550/arXiv.2307.08835}

\bibitem[{{Trinca} {et~al.}(2023){Trinca}, {Schneider}, {Valiante}, {Graziani}, {Ferrotti}, {Omukai}, \& {Chon}}]{Trinca_2023}
{Trinca}, A., {Schneider}, R., {Valiante}, R., {et~al.} 2023, arXiv e-prints, arXiv:2305.04944, \dodoi{10.48550/arXiv.2305.04944}

\bibitem[{{Vijayan} {et~al.}(2021){Vijayan}, {Lovell}, {Wilkins}, {Thomas}, {Barnes}, {Irodotou}, {Kuusisto}, \& {Roper}}]{Vijayan_2021}
{Vijayan}, A.~P., {Lovell}, C.~C., {Wilkins}, S.~M., {et~al.} 2021, \mnras, 501, 3289, \dodoi{10.1093/mnras/staa3715}

\bibitem[{{Vogelsberger} {et~al.}(2020){Vogelsberger}, {Nelson}, {Pillepich}, {Shen}, {Marinacci}, {Springel}, {Pakmor}, {Tacchella}, {Weinberger}, {Torrey}, \& {Hernquist}}]{Vogelsberger_2020}
{Vogelsberger}, M., {Nelson}, D., {Pillepich}, A., {et~al.} 2020, \mnras, 492, 5167, \dodoi{10.1093/mnras/staa137}

\bibitem[{{Wetzel} \& {Garrison-Kimmel}(2020)}]{2020ascl.soft02015W}
{Wetzel}, A., \& {Garrison-Kimmel}, S. 2020, {GizmoAnalysis: Read and analyze Gizmo simulations}, Astrophysics Source Code Library, record ascl:2002.015.
\newblock \doeprint{2002.015}

\bibitem[{{Wetzel} {et~al.}(2023){Wetzel}, {Hayward}, {Sanderson}, {Ma}, {Angl{\'e}s-Alc{\'a}zar}, {Feldmann}, {Chan}, {El-Badry}, {Wheeler}, {Garrison-Kimmel}, {Nikakhtar}, {Panithanpaisal}, {Arora}, {Gurvich}, {Samuel}, {Sameie}, {Pandya}, {Hafen}, {Hummels}, {Loebman}, {Boylan-Kolchin}, {Bullock}, {Faucher-Gigu{\`e}re}, {Kere{\v{s}}}, {Quataert}, \& {Hopkins}}]{Wetzel_2023}
{Wetzel}, A., {Hayward}, C.~C., {Sanderson}, R.~E., {et~al.} 2023, \apjs, 265, 44, \dodoi{10.3847/1538-4365/acb99a}

\bibitem[{{Wetzel} {et~al.}(2016){Wetzel}, {Hopkins}, {Kim}, {Faucher-Gigu{\`e}re}, {Kere{\v{s}}}, \& {Quataert}}]{Wetzel_2016}
{Wetzel}, A.~R., {Hopkins}, P.~F., {Kim}, J.-h., {et~al.} 2016, \apjl, 827, L23, \dodoi{10.3847/2041-8205/827/2/L23}

\bibitem[{{Wilkins} {et~al.}(2023){Wilkins}, {Vijayan}, {Lovell}, {Roper}, {Irodotou}, {Caruana}, {Seeyave}, {Kuusisto}, {Thomas}, \& {Parris}}]{Wilkins_2023}
{Wilkins}, S.~M., {Vijayan}, A.~P., {Lovell}, C.~C., {et~al.} 2023, \mnras, 519, 3118, \dodoi{10.1093/mnras/stac3280}

\bibitem[{{Witstok} {et~al.}(2023){Witstok}, {Shivaei}, {Smit}, {Maiolino}, {Carniani}, {Curtis-Lake}, {Ferruit}, {Arribas}, {Bunker}, {Cameron}, {Charlot}, {Chevallard}, {Curti}, {de Graaff}, {D'Eugenio}, {Giardino}, {Looser}, {Rawle}, {Rodr{\'\i}guez del Pino}, {Willott}, {Alberts}, {Baker}, {Boyett}, {Egami}, {Eisenstein}, {Endsley}, {Hainline}, {Ji}, {Johnson}, {Kumari}, {Lyu}, {Nelson}, {Perna}, {Rieke}, {Robertson}, {Sandles}, {Saxena}, {Scholtz}, {Sun}, {Tacchella}, {Williams}, \& {Willmer}}]{Witstok_2023}
{Witstok}, J., {Shivaei}, I., {Smit}, R., {et~al.} 2023, arXiv e-prints, arXiv:2302.05468, \dodoi{10.48550/arXiv.2302.05468}

\bibitem[{{Yan} {et~al.}(2023){Yan}, {Ma}, {Ling}, {Cheng}, \& {Huang}}]{Yan_2023}
{Yan}, H., {Ma}, Z., {Ling}, C., {Cheng}, C., \& {Huang}, J.-S. 2023, \apjl, 942, L9, \dodoi{10.3847/2041-8213/aca80c}

\bibitem[{{Yung} {et~al.}(2023){Yung}, {Somerville}, {Finkelstein}, {Wilkins}, \& {Gardner}}]{Yung_2023}
{Yung}, L.~Y.~A., {Somerville}, R.~S., {Finkelstein}, S.~L., {Wilkins}, S.~M., \& {Gardner}, J.~P. 2023, arXiv e-prints, arXiv:2304.04348, \dodoi{10.48550/arXiv.2304.04348}

\bibitem[{{Zavala} {et~al.}(2023){Zavala}, {Buat}, {Casey}, {Finkelstein}, {Burgarella}, {Bagley}, {Ciesla}, {Daddi}, {Dickinson}, {Ferguson}, {Franco}, {Jim{\'e}nez-Andrade}, {Kartaltepe}, {Koekemoer}, {Le Bail}, {Murphy}, {Papovich}, {Tacchella}, {Wilkins}, {Aretxaga}, {Behroozi}, {Champagne}, {Fontana}, {Giavalisco}, {Grazian}, {Grogin}, {Kewley}, {Kocevski}, {Kirkpatrick}, {Lotz}, {Pentericci}, {P{\'e}rez-Gonz{\'a}lez}, {Pirzkal}, {Ravindranath}, {Somerville}, {Trump}, {Yang}, {Yung}, {Almaini}, {Amor{\'\i}n}, {Annunziatella}, {Haro}, {Backhaus}, {Barro}, {Bell}, {Bhatawdekar}, {Bisigello}, {Buitrago}, {Calabr{\`o}}, {Castellano}, {Ch{\'a}vez Ortiz}, {Chworowsky}, {Cleri}, {Cohen}, {Cole}, {Cooke}, {Cooper}, {Cooray}, {Costantin}, {Cox}, {Croton}, {Dav{\'e}}, {de La Vega}, {Dekel}, {Elbaz}, {Estrada-Carpenter}, {Fern{\'a}ndez}, {Finkelstein}, {Freundlich}, {Fujimoto}, {Garc{\'\i}a-Argum{\'a}nez}, {Gardner}, {Gawiser}, {G{\'o}mez-Guijarro}, {Guo}, {Hamilton}, {Hathi}, {Holwerda}, {Hirschmann},
  {Huertas-Company}, {Hutchison}, {Iyer}, {Jaskot}, {Jha}, {Jogee}, {Juneau}, {Jung}, {Kassin}, {Kurczynski}, {Larson}, {Leung}, {Long}, {Lucas}, {Magnelli}, {Mantha}, {Matharu}, {McGrath}, {McIntosh}, {Medrano}, {Merlin}, {Mobasher}, {Morales}, {Newman}, {Nicholls}, {Pandya}, {Rafelski}, {Ronayne}, {Rose}, {Ryan}, {Santini}, {Seill{\'e}}, {Shah}, {Shen}, {Simons}, {Snyder}, {Stanway}, {Straughn}, {Teplitz}, {Vanderhoof}, {Vega-Ferrero}, {Wang}, {Weiner}, {Willmer}, {Wuyts}, \& {CEERS Team}}]{Zavala_2023}
{Zavala}, J.~A., {Buat}, V., {Casey}, C.~M., {et~al.} 2023, \apjl, 943, L9, \dodoi{10.3847/2041-8213/acacfe}

\end{thebibliography}
\bibliographystyle{aasjournal}



\end{document}